\documentclass[reprint, 10pt, amsmath, amssymb, floatfix, aps, prd,  superscriptaddress, nofootinbib, nobibnotes,twocolumn]{revtex4-1}

\pdfoutput=1
\usepackage{graphicx}
\graphicspath{{figures/}}
\usepackage[utf8]{inputenc}
\usepackage{amsmath}
\usepackage{amsfonts}
\usepackage{amssymb}
\usepackage{multirow}
\usepackage{tcolorbox}
\usepackage{CJKutf8}
\usepackage{color}
\usepackage[colorlinks=true, linkcolor=red, citecolor=blue]{hyperref}
\usepackage{orcidlink}
\usepackage[capitalise]{cleveref}
\usepackage{acro}
\usepackage{adjustbox}
\usepackage{array}
\usepackage{natbib}
\usepackage{dblfnote}
\DFNalwaysdouble
\usepackage{slashed}
\usepackage{dcolumn}
\usepackage{hyperref}
\usepackage{geometry}
\usepackage{subfig} 
\usepackage{overpic}
\usepackage{inputenc}
\usepackage{caption}
\usepackage{tikz}
\usepackage{tabularx}
\usepackage[compat=1.1.0]{tikz-feynhand}
\def\({\left(}
\def\){\right)}
\def\[{\left[}
\def\]{\right]}
\def\be{\begin{eqnarray}}
\def\ee{\end{eqnarray}}

\DeclareAcronym{GW}{
  short = GW ,
  long = gravitational wave ,
  short-plural = s 
}
\DeclareAcronym{LIGO}{
  short = LIGO ,
  long = Laser Interferometer Gravitational-wave Observatory ,
  short-plural = 
}
\DeclareAcronym{LISA}{
  short = LISA ,
  long = Laser Interferometer Space Antenna ,
  short-plural =  
}
\DeclareAcronym{SKA}{
  short = SKA ,
  long = Square Kilometre Array ,
  short-plural =  
}

\DeclareAcronym{SNR}{
	short = SNR ,
	long = signal-to-noise ratio ,
	short-plural = 
}

\DeclareAcronym{PTA}{
	short = PTA ,
	long = pulsar timing array ,
	short-plural = 
}

\DeclareAcronym{FLRW}{
  short = FLRW ,
  long = Friedmann-Lemaitre-Robertson-Walker ,
  short-plural =  
}

\DeclareAcronym{SIGW}{
	short = SIGW ,
	long = scalar induced gravitational wave ,
	short-plural =  s
}

\DeclareAcronym{TSIGW}{
	short = TSIGW ,
	long = tensor-scalar induced gravitational wave ,
	short-plural =  s
}

\DeclareAcronym{1PI}{
	short = 1PI ,
	long = one-particle irreducible  ,
	short-plural =  
}

\DeclareAcronym{1PR}{
	short = 1PR ,
	long = one-particle reducible  ,
	short-plural =  
}

\DeclareAcronym{PBH}{
	short = PBH ,
	long = primordial black hole ,
	short-plural =  s
}

\DeclareAcronym{CMB}{
	short = CMB ,
	long = cosmic microwave background ,
	short-plural =  
}
\DeclareAcronym{DM}{
	short = DM ,
	long = dark matter ,
	short-plural =  
}

\DeclareAcronym{SGWB}{
	short = SGWB ,
	long = stochastic gravitational	wave background ,
	short-plural =  s
}

\DeclareAcronym{LSS}{
	short = LSS ,
	long = large scale structure ,
	short-plural =  
}

\DeclareAcronym{RD}{
	short = RD ,
	long = radiation-dominated ,
	short-plural =  
}

\DeclareAcronym{SM}{
	short = SM ,
	long = supplementary material ,
	short-plural =  
}

\DeclareAcronym{CP}{
	short = CP ,
	long = cosmological perturbation ,
	short-plural =  s
}

\DeclareAcronym{QFT}{
	short = QFT ,
	long = quantum field theory ,
	short-plural =  
}

\begin{document}

\title{ Induced gravitational waves for arbitrary higher orders: vertex rules and loop diagrams in cosmological perturbation theory}

\author{Jing-Zhi Zhou\orcidlink{0000-0003-2792-3182}} 
\email{zhoujingzhi@tju.edu.cn} 
\affiliation{Center for Joint Quantum Studies and Department of Physics, School of Science, Tianjin University, Tianjin 300350, China} 

\author{Yu-Ting Kuang\orcidlink{0000-0002-7431-4454}} 
\affiliation{Institute of High Energy Physics, Chinese Academy of Sciences, Beijing 100049, China} 
\affiliation{University of Chinese Academy of Sciences, Beijing 100049, China} 

\author{Di Wu\orcidlink{0000-0001-7309-574X}} 
\affiliation{Institute of High Energy Physics, Chinese Academy of Sciences, Beijing 100049, China} 
\affiliation{University of Chinese Academy of Sciences, Beijing 100049, China} 

\author{ H. L\"u\orcidlink{0000-0001-7100-2466}} 
\affiliation{Center for Joint Quantum Studies and Department of Physics, School of Science, Tianjin University, Tianjin 300350, China}
\affiliation{Joint School of National University of Singapore and Tianjin University, International Campus of Tianjin University, Binhai New City, Fuzhou 350207, China} 

\author{Zhe Chang\orcidlink{0000-0002-9720-803X}} 
\affiliation{Institute of High Energy Physics, Chinese Academy of Sciences, Beijing 100049, China} 
\affiliation{University of Chinese Academy of Sciences, Beijing 100049, China}

\begin{abstract}
Gravitational waves induced by primordial perturbations serve as crucial probes for studying the early universe, providing  a significant window into potential new physics during cosmic evolution. Due to the potentially large amplitudes of primordial perturbations on small scales, the contributions of high-order \acp{CP} are highly significant. We propose a vertex approach applicable to the study of induced gravitational waves for arbitrary higher orders. Using the vertex approach and tree diagrams, we can directly derive the explicit expressions of higher-order induced gravitational waves without involving the complex and lengthy calculations of higher-order \acp{CP}. Correlations between different tree diagrams correspond to the loop diagrams of two-point correlation functions of induced gravitational waves. Our investigation reveals that \ac{1PR} diagrams impact \acp{TSIGW} while leaving \acp{SIGW} unaffected.
\end{abstract}

\maketitle

\acresetall

\section{Introduction}\label{sec:1.0}
Our universe originated from primordial perturbations generated during the inflationary epoch \cite{Bartolo:2004if,Lyth:1998xn,Novello:2008ra}. These perturbations, serving as the initial conditions for cosmic evolution, encode crucial physical insights into the early universe. By leveraging current cosmological observations—such as \ac{CMB}, \ac{LSS}, and \acp{PBH}—we can discern the physical properties of primordial perturbations across various scales \cite{Planck:2018vyg,Planck:2018nkj,Carr:2016drx,Carr:2020gox}. In June 2023, several international \ac{PTA} collaborations, such as NANOGrav \cite{NANOGrav:2023gor}, PPTA \cite{Reardon:2023gzh}, EPTA \cite{EPTA:2023fyk}, and the CPTA \cite{Xu:2023wog}, provided evidence for the existence of a \ac{SGWB} in the nHz frequency range. This breakthrough provides a fresh avenue for investigating potential new physics during the early universe and characterizing the physical properties of small-scale primordial perturbations. Specifically, primordial perturbations generated during the inflationary epoch, upon re-entering the horizon after inflation, can induce \acp{GW} with discernible effects \cite{Domenech:2021ztg}.  Current \ac{PTA} observational data suggest that induced \acp{GW} are among the most likely sources contributing to the \acp{SGWB} \cite{NANOGrav:2023hvm}. The imprint of potential new physics during inflation and the nature of small-scale primordial perturbations are encoded in today’s PTA observations   \cite{Wang:2023ost,Choudhury:2023fwk,Ellis:2023oxs,Balaji:2023ehk,You:2023rmn}.

The investigation into \acp{SIGW} on small scales is inspired by research on the impact of second-order tensor and vector perturbations induced by first-order scalars on \acp{CMB} polarization \cite{Mollerach:2003nq}. Building on previous research on second-order perturbations at the \acp{CMB} scale, Refs.~\cite{Ananda:2006af,Osano:2006ew,Baumann:2007zm} studied the second-order \acp{GW} induced by large-amplitude primordial scalar perturbations on small scales and calculated the corresponding energy density spectrum. In the past decade, researches on second-order \acp{SIGW} has been extended to \acp{PBH} \cite{Wang:2019kaf,Byrnes:2018txb,Inomata:2020lmk,Ballesteros:2020qam,Lin:2020goi,Chen:2019xse,Cai:2019elf,Cai:2019jah,Ando:2018qdb,Di:2017ndc,Gao:2021vxb,Changa:2022trj,Zhou:2020kkf,Cai:2021wzd,Domenech:2024cjn}, gauge issue \cite{Hwang:2017oxa,Yuan:2019fwv,Inomata:2019yww,DeLuca:2019ufz,Domenech:2020xin,Chang:2020tji,Ali:2020sfw,Lu:2020diy,Tomikawa:2019tvi,Gurian:2021rfv,Uggla:2018fiy,Ali:2023moi}, primordial non-Gaussianity \cite{Cai:2018dig,Atal:2021jyo,Zhang:2020uek,Yuan:2020iwf,Davies:2021loj,Rezazadeh:2021clf,Kristiano:2021urj,Bartolo:2018qqn,Adshead:2021hnm,Li:2023qua,Li:2023xtl,Garcia-Saenz:2022tzu,Li:2024zwx,Perna:2024ehx}, different epochs of the Universe \cite{Kohri:2018awv,Papanikolaou:2020qtd,Domenech:2020kqm,Domenech:2019quo,Inomata:2019zqy,Inomata:2019ivs,Assadullahi:2009nf,Witkowski:2021raz,Dalianis:2020gup,Hajkarim:2019nbx,Bernal:2019lpc,Das:2021wad,Haque:2021dha,Domenech:2020ssp,Domenech:2021and,Liu:2023pau}, and damping effect  \cite{Mangilli:2008bw,Saga:2014jca,Zhang:2022dgx,Yuan:2023ofl}.

As a measurable quantity within \ac{PTA} experiments, the energy density spectrum of \acp{SIGW} can be expressed as  $\Omega_{\mathrm{GW}}\left( k\right)=A^2_{\zeta}\Omega^{(2)}_{\mathrm{GW}}\left( k\right)+A^3_{\zeta}\Omega^{(3)}_{\mathrm{GW}}\left( k\right)+\cdots$, where the perturbation expansion parameter $A_{\zeta}$ represents the amplitude of the power spectrum of primordial curvature perturbation. While previous studies mainly focused on second-order \acp{SIGW}, recent research emphasizes the impact of third-order \acp{SIGW} on current \ac{PTA} observations \cite{Chang:2023vjk,Chang:2023aba,Zhou:2021vcw,Chang:2022nzu,Yuan:2019udt,Wang:2023sij}. Disregarding higher-order effects will inevitably lead to deviations in the primordial perturbation parameters derived from the analysing of \ac{PTA} data. Consequently, there is a compelling physical motivation to meticulously compute or rigorously estimate the higher-order effects arising from induced \acp{GW}. However, computing higher-order induced \acp{GW} remains an arduous endeavor, requiring rigorous derivation and solution of higher-order \ac{CP} equations in \ac{FLRW} spacetime, followed by the evaluation of relevant two-point correlation functions \cite{Malik:2008im,Ma:1995ey}. More precisely, in the traditional
method, the calculation of higher-order induced \acp{GW} involves three main steps:
\tcbset{colback=gray!20, colframe=gray!20, boxrule=0.5mm, arc=0mm, auto outer arc, width=\linewidth} 
\begin{tcolorbox} 
\noindent
\textbf{ 1, In \ac{FLRW} spacetime, perturbatively expand the Einstein field equations step by step, extracting the \acp{CP} equations at each order.}

\

\noindent
\textbf{ 2, Solve the \acp{CP} equations step by step, starting from the first-order \acp{CP}.}

\

\noindent
\textbf{ 3, Using the explicit expression of the $n$-th order induced \acp{GW} obtained, compute the two-point correlation function and the corresponding power spectrum for the $n$-th order induced \acp{GW}.}
\end{tcolorbox}
\noindent
Due to the complexity of the calculation process, it is almost impossible to study higher-order induced gravitational waves using the aforementioned traditional methods. For \acp{CP} beyond the third order, deriving the \ac{CP} equations in the initial step is nearly unachievable.

In this study, we present a systematic framework for investigating higher-order induced \acp{GW}, introducing both vertex and loop diagram approaches. These approaches empower us to explore arbitrary higher-order scenarios. Specifically, the calculation of higher-order induced \acp{GW} using vertex and loop diagram approaches can be divided into the following steps:
\tcbset{colback=gray!20, colframe=gray!20, boxrule=0.5mm, arc=0mm, auto outer arc, width=\linewidth} 
\begin{tcolorbox} 
\noindent
\textbf{ 1, Based on the vertex rules, generate all tree diagrams corresponding to the $n$-th order induced \acp{GW}, and then sum the expressions of these tree diagrams to obtain the specific expression of the $n$-th order induced \acp{GW}.}

\

\noindent
\textbf{ 2, Identify all possible loop diagrams for the $n$-th order induced \acp{GW} and sum these loop diagrams to derive the energy density spectrum of the n-th order induced \acp{GW}.}
\end{tcolorbox}
\noindent
The first step of the aforementioned calculation process involves the vertex rules and tree diagrams. Using the vertex approach, we can directly skip the first two parts of the calculation in the traditional method. The explicit expression for higher-order induced \acp{GW} can be derived with pen and paper, without involving thousands of terms from complex and lengthy calculations of higher-order \acp{CP}. The second step in the calculation process involves the loop diagram structure of the induced \acp{GW}. In the loop diagram approach, we can use different types of loop diagrams to calculate the two-point correlation function of induced \acp{GW}. A comprehensive examination of various loop diagram structures allows for more efficient calculation of momentum integrals within the power spectrum of higher-order induced \acp{GW}. In this paper, we systematically analyze the differences between \acp{TSIGW} and \acp{SIGW} in the context of \ac{1PR} diagrams. Our investigation reveals that \ac{1PR} diagrams affect \acp{TSIGW} while leaving \acp{SIGW} unaffected \cite{Chang:2022vlv,Bari:2023rcw,Chen:2022dah}. 

This paper is organized as follows. In Sec.~\ref{sec:2.0}, we provide a concise review of the computation of high-order \acp{SIGW}, highlighting potential redundancies within the calculations. In Sec.~\ref{sec:3.0}, we attempt to eliminate these redundancies in the calculation of higher-order \acp{CP} and provide the specific form of the vertex rules of induced \acp{GW}. In Sec.~\ref{sec:4.0}, we explore the general structure of loop diagrams used to calculate the power spectrum of induced \acp{GW} and investigate the differences between \acp{SIGW} and \acp{TSIGW} at the \ac{1PR} diagram level. Finally, we summarize our results and give some discussions in Sec.~\ref{sec:5.0}.

\section{Equations of motion}\label{sec:2.0}
In this section, we briefly review the calculation methods for induced \acp{GW} in traditional approaches and attempt to eliminate redundancies to simplify the calculation of higher-order \acp{CP}.

\subsection{Review of SIGWs}\label{sec:2.1}
In traditional methods, to calculate the energy density spectrum of higher-order \acp{SIGW}, the first step is to compute higher-order \acp{CP} and derive the equation of motion for $n$-th order \acp{SIGW}. For instance, the third-order perturbation of the \ac{FLRW} metric in Newtonian gauge can be written as
\begin{equation}\label{eq:3ods}
	\begin{aligned}
		&\mathrm{d}s^{2}=a^{2}\left(-\left(1+2 \phi^{(1)}+ \phi^{(2)}\right) \mathrm{d} \eta^{2}+ V_i^{(2)} \mathrm{d} \eta \mathrm{d} x^{i}+\right. \\
		&\left.\left(\left(1-2 \psi^{(1)}- \psi^{(2)}\right) \delta_{i j}+\frac{1}{2} h_{i j}^{(2)}+\frac{1}{6} h_{i j}^{(3)}\right)\mathrm{d} x^{i} \mathrm{d} x^{j}\right) \ ,
	\end{aligned}
\end{equation}
where $\phi^{(n)}$ and $\psi^{(n)}$$\left( n=1,2 \right)$ are first order and second order scalar perturbations. $h^{(n)}_{ij}$$\left( n=2,3 \right)$ are second-order and third-order tensor perturbations. $V_i^{(2)} $ is second order vector perturbation. Since we focus on \acp{SIGW}, we have neglected first-order tensor and vector perturbations. To derive the equation of motion for third-order \acp{SIGW}, we need to substitute Eq.~(\ref{eq:3ods}) and the energy-momentum tensor of an ideal fluid into Einstein's field equation $G_{\mu\nu}=\kappa T_{\mu\nu}$, and extract the equations of motion for each order of \acp{CP}. Typically, we utilize the \texttt{xPand} package to aid in the calculations of higher-order \acp{CP} \cite{Pitrou:2013hga}. When dealing with third-order \acp{CP}, the perturbation equations contain more than two thousand terms.

The complexity of higher-order \ac{CP} equations makes it nearly impossible to compute perturbations beyond the third order. Therefore, with conventional methods, deriving the motion equations for higher-order induced \acp{GW} is almost infeasible, let alone solving these equations and computing the power spectrum of induced \acp{GW}. To systematically investigate or precisely estimate the effects of higher-order \acp{CP} on small scales, it is crucial to simplify the calculations of the higher-order \acp{CP}.

As shown in Eq.~(\ref{eq:3ods}), before substituting  the metric perturbations into Einstein's field equations, we have already performed a perturbative expansion of the \acp{CP}. For example, for scalar perturbations, $\phi=\phi^{(1)}+\frac{1}{2}\phi^{(2)}+\cdots$. This approach is evidently straightforward and logical. In traditional methods, we first need to determine the order of the induced \acp{GW} to be calculated, then expand the metric perturbation to the corresponding order and substitute them into Einstein's field equations. However, as we will demonstrate, this method also contributes to redundancy in the calculation of higher-order \acp{CP}. The higher the order of \acp{CP} to be computed, the more cumbersome the expressions for metric perturbations become, and the more complex the calculation of \acp{CP}.

\subsection{Redundancy}\label{sec:2.2}
The complexity of higher-order \acp{CP} mainly arises from the intricate perturbative expansions of metric perturbations. Consequently, to streamline the calculation of higher-order \acp{CP}, it is essential to examine the relationship between different forms of metric perturbation expansions and their corresponding equations of motion.  More precisely, the perturbed metric in the \ac{FLRW} spacetime with
Newtonian gauge takes the form
\begin{equation}\label{eq:FL}
	\begin{aligned}
		\mathrm{d}s^{2}=a^{2}& \left(-\left(1+2 \phi \right) \mathrm{d} \eta^{2}+ 2V_i \mathrm{d} \eta \mathrm{d} x^{i}\right. \\
		&~\left.+\left(\left(1-2 \psi \right) \delta_{i j}+h_{ij}\right)\mathrm{d} x^{i} \mathrm{d} x^{j}\right) \ .
	\end{aligned}
\end{equation}
The \acp{CP} $A$ (where $A =\phi,\psi,V_i$, and $h_{ij}$ in Eq.~(\ref{eq:FL})) can be expressed as $A=\sum_{n=1}^{\infty}\frac{1}{n!}A^{(n)}$. Within the specified metric perturbation framework, we can conveniently investigate the influence of different lower-order perturbations on higher-order induced \acp{GW}. For instance, considering second-order \acp{GW} $h^{(2)}_{ij}$ induced by first-order scalar perturbation $\phi^{(1)}$, we set $\phi=\phi^{(1)}$, $h_{ij}=\frac{1}{2}h^{(2)}_{ij}$, and $\psi=V_i=0$. Namely, 
\begin{equation}\label{eq:FL2}
	\begin{aligned}
		\mathrm{d}s^{2}=a^{2}& \left(-\left(1+2 \phi^{(1)} \right) \mathrm{d} \eta^{2}\right. \\
		&~~~~\left.+\left( \delta_{i j}+\frac{1}{2}h^{(2)}_{ij}\right)\mathrm{d} x^{i} \mathrm{d} x^{j}\right) \ .
	\end{aligned}
\end{equation}
By substituting the perturbed metric in Eq.~(\ref{eq:FL2}) into the Einstein's field equation and simplifying, we derive the equation of motion for second-order \acp{GW} induced by first-order scalar perturbation $\phi^{(1)}$  during the \ac{RD} era
\begin{eqnarray}
        h_{lm}^{(2)''} +2 \mathcal{H}  h_{lm}^{(2)'}-\Delta h_{lm}^{(2)} =-8\Lambda_{lm}^{rs} \phi^{(1)}\partial_r\partial_s\phi^{(1)}  , \label{eq:112}
\end{eqnarray}
where $\Lambda_{rs}^{l m}=\mathcal{T}_{r}^{l} \mathcal{T}_{s}^{m}-\frac{1}{2} \mathcal{T}_{rs} \mathcal{T}^{l m}$ is the transverse and traceless decomposition operator, and $\mathcal{T}_{r}^{l}$ is defined as $\mathcal{T}_{r}^{l}=\delta_{r}^{l}-\partial^{l} \Delta^{-1} \partial_{r}$ \cite{Chang:2020tji}.

Similarly, for fourth-order \acp{GW} $h^{(4)}_{ij}$ induced by second-order scalar perturbation $\phi^{(2)}$, we set $\phi=\frac{1}{2}\phi^{(2)}$, $h_{ij}=\frac{1}{4 !}h^{(4)}_{ij}$, and $\psi=V_i=0$.  The perturbed metric in Eq.~(\ref{eq:FL}) can be expressed as
\begin{equation}\label{eq:FL4}
	\begin{aligned}
		\mathrm{d}s^{2}=a^{2}& \left(-\left(1+ \phi^{(2)} \right) \mathrm{d} \eta^{2}\right. \\
		&~~~~\left.+\left( \delta_{i j}+\frac{1}{4!}h^{(4)}_{ij}\right)\mathrm{d} x^{i} \mathrm{d} x^{j}\right) \ .
	\end{aligned}
\end{equation}
 Interestingly, comparing the perturbed metrics between Eq.~(\ref{eq:FL2}) and Eq.~(\ref{eq:FL4}) reveals that their forms are entirely identical. The transformation between these two perturbed metrics can be achieved through a straightforward variable replacement: $\phi^{(1)}\to \frac{1}{2}\phi^{(2)}$ and $\frac{1}{2}h^{(2)}_{ij}\to \frac{1}{4 !}h^{(4)}_{ij}$.  By applying the variable replacement to Eq.~(\ref{eq:112}), we obtain 
\begin{eqnarray}
        h_{lm}^{(4)''} +2 \mathcal{H}  h_{lm}^{(4)'}-\Delta h_{lm}^{(4)} =-48\Lambda_{lm}^{rs} \phi^{(2)}\partial_r\partial_s\phi^{(2)} . \label{eq:244}
\end{eqnarray}
This equation is consistent with the direct computational outcome from higher-order \ac{CP} theory. Consequently, we reach a crucial conclusion: the equation of motion for second-order \acp{GW} induced by first-order scalar perturbation $\phi^{(1)}$ shares the same form as that for fourth-order \acp{GW} induced by second-order scalar perturbation $\phi^{(2)}$, except for the Taylor expansion coefficients. Since the perturbed metrics in Eq.~(\ref{eq:FL2}) and Eq.~(\ref{eq:FL4}) have exactly the same form, it is not surprising that they yield perturbation equations with identical forms.

The above discussion inspires us that the form of the source term in the equation of motion of induced \acp{GW} is related only to the type and number of lower-order \acp{CP} in the source term, and not to the order of these \acp{CP}. As discussed earlier, in traditional methods, directly performing a perturbative expansion of \acp{CP} introduces a substantial amount of redundancy. Thus, if we directly substitute the metric perturbation in Eq.~(\ref{eq:FL}) into the Einstein field equations, and then carry out a perturbative expansion on various \acp{CP} after deriving the motion equation of the induced \acp{GW}, this redundancy can be effectively eliminated. After eliminating these redundancies, the calculation  of higher-order \acp{CP} becomes feasible. Building upon this insight, we will present the vertex approach for studying higher-order \acp{GW}, streamlining and enhancing the efficiency of the research on higher-order effects.

\section{Vertex rules and Tree diagrams}\label{sec:3.0}
In Sec.~\ref{sec:2.0}, we investigated the origins of complexity and redundancy in computing higher-order \acp{CP}. To eliminate redundancy in the calculation of higher-order \acp{CP}, we can derive the higher-order \ac{CP} equations directly without initially performing a perturbative expansion for the \acp{CP}. After deriving the \ac{CP} equations, we then perform the perturbative expansion for these \acp{CP} and simplify the resulting equations. By utilizing this method, we find that the source terms in the equations of motion for higher-order induced \acp{GW} depend solely on the types and quantities of lower-order \acp{CP} in the source terms, rather than their specific orders. To efficiently compute higher-order induced \acp{GW}, we summarize the computation methods into several vertex rules. Based on these vertex rules and the corresponding tree diagrams, we can directly write the explicit expressions for higher-order induced \acp{GW} without deriving and solving complex higher-order \ac{CP} equations. 

\subsection{Vertex rules}\label{sec:3.1}
Since the calculation of $n+1$-th order induced \acp{GW} involves all lower-order \acp{CP}, the vertex rules must be applicable to all types of \acp{CP}. In Newtonian gauge, the equations of motion for the four types of \acp{CP} in Eq.~(\ref{eq:FL}) during the \ac{RD} era can be expressed as \cite{Lu:2008ju,Chang:2022aqk}
\begin{eqnarray}
  &\psi^{(n)''} +3\mathcal{H}  \psi^{(n)'} -\frac{5}{6} \Delta \psi^{(n)}+\mathcal{H}  \phi^{(n)'} +\frac{1}{2} \Delta \phi^{(n)} \nonumber\\
		&~~~~=-\frac{n!}{4} \mathcal{T}^{rs} \mathcal{S}^{(n)}_{rs}  \ , \label{eq:2eS2}\\
  &\psi^{(n)} -\phi^{(n)} =-n! \Delta^{-1}\left(\partial^{r} \Delta^{-1} \partial^{s}\right.\nonumber\\
&~~~~~\left.-\frac{1}{2} \mathcal{T}^{rs}\right) \mathcal{S}^{(n)}_{rs}  \ , \label{eq:2eS1}
\end{eqnarray}
\begin{eqnarray}
        &V_{l}^{(n)'} +2 \mathcal{H} V_{l}^{(n)}  =2n! \Delta^{-1} \mathcal{T}_{l}^{r} \partial^{s} \mathcal{S}^{(n)}_{rs}  \ , \label{eq:2eV}
        \end{eqnarray}
        \begin{eqnarray}
        &h_{lm}^{(n)''} +2 \mathcal{H}  h_{lm}^{(n)'} -\Delta h_{lm}^{(n)} =-2n! \Lambda_{lm}^{rs} \mathcal{S}^{(n)}_{rs}  \ , \label{eq:2eT}
\end{eqnarray}
where $\mathcal{H}=a'/a=1/\eta$ is the conformal Hubble
parameter. Superscript $n$ represents the order of \acp{CP}, and $\mathcal{S}^{(n)}_{rs}$ denotes the source term composed of lower-order \acp{CP}. In Eq.~(\ref{eq:2eS2})--Eq.~(\ref{eq:2eT}),  we have assumed $p^{(n)}=\frac{1}{3}\rho^{(n)}$, where $p^{(n)}$ and $\rho^{(n)}$ are $n$-th order pressure and density perturbations, respectively. Eq.~(\ref{eq:2eS2})--Eq.~(\ref{eq:2eT}) can be studied in terms of vertex diagrams.  We summarize the physical information contained in the $n$-branch vertex diagrams as follows: 
\begin{figure}[htbp]
    \centering
    \includegraphics[width=0.6\columnwidth]{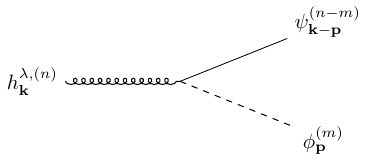}
\caption{\label{fig:1} The 2-branch vertex diagram of higher-order \acp{GW} induced by the product of $\phi$ and $\psi$. }
\end{figure}

\noindent
\textbf{a,  Green’s function and decomposed operators:} the curve on the left side of the vertex diagram in Fig.~\ref{fig:1} corresponds to the $n$-th order \ac{CP} that needs to be solved. The dynamical terms and decomposition operators in $n$-th order \ac{CP} equations depend exclusively on the type of perturbation, resulting in the same Green’s function for perturbations of the same type. Once the curve on the left side of the vertex diagram is determined, the Green’s function of the perturbation equation and the corresponding decomposition operator are fully specified. The formal expression for the higher-order induced \acp{GW} corresponding to the vertex diagram in Fig.~\ref{fig:1} is given by
\begin{eqnarray}
        h^{\lambda,(n)}_{\mathbf{k},\psi\phi}(\eta)&=&-\sum_{A=\psi,\phi}\frac{2n!}{k} \int^{\eta}_0 \mathrm{d} \tilde{\eta}  \sin (k \eta-k \tilde{\eta}) \frac{\tilde{\eta}}{\eta}  \nonumber\\
        &\times& \varepsilon^{\lambda, ij}(\mathbf{k}) S_{ij,\psi\phi}^{(n)}(\mathbf{k},\tilde{\eta}) \ , \label{eq:hv}
\end{eqnarray}
where the symbol $h^{\lambda,(n)}_{\mathbf{k},\psi\phi}$ represents the higher-order \acp{SIGW} sourced by the product of $\phi$ and $\psi$ \cite{Ellis:2016jkw}. The formal expressions of $\phi^{(n)}_{\mathbf{k}}$, $\psi^{(n)}_{\mathbf{k}}$, and $V^{\lambda,(n)}_{\mathbf{k}}$ can be found in Refs.~\cite{Chang:2022dhh,Inomata:2020cck}.

\noindent
\textbf{b, Source term:} the branches on the right side of the vertex diagram correspond to the source terms formed by lower-order \acp{CP}. As mentioned in the previous section, higher-order source terms can be constructed by replacing variables in lower-order source terms. When the number and types of branches on the right side are completely determined, the form of the source terms is also fully determined. The explicit expression for the source term corresponding to Fig.~\ref{fig:1} is as follows
\begin{eqnarray}
     & &S_{ij,\psi\phi}^{(n)}(\mathbf{k},\eta)=\frac{1}{m!(n-m)!}\int \frac{\mathrm{d}^3 p}{(2 \pi)^{3 / 2}}  p_i p_j \nonumber\\
     & &~~~~~~~~~~~~~~~\times~\frac{2}{\mathcal{H}}\psi^{(n-m)'}_{\mathbf{k}-\mathbf{p}}\phi^{(m)}_{\mathbf{p}}  \ , \label{eq:Sv2}
\end{eqnarray}
where the source term has been simplified using the property of the transverse and traceless operator $\Lambda^{rs}_{ij}~\partial_i\phi\partial_j\phi=-\Lambda^{rs}_{ij}~\phi\partial_i\partial_j\phi$ .

\noindent
\textbf{c,  Momentum integrals:} in the case of a vertex diagram with $n$ branches on the right-hand side, the associated source term is formed by multiplying $n$ lower-order perturbations. Consequently, the expression for an $n$-branch vertex diagram will     include $n-1$ momentum integrals.

In the Appendix.~\ref{sec:A}, we derive the explicit expressions for $n$-th order perturbations of the Einstein's field equation  during the \ac{RD} era in Newtonian gauge. The source terms of induced \acp{GW} associated with all types of 2-branch vertex diagrams are given in Appendix.~\ref{sec:B}.

\subsection{Tree diagrams}\label{sec:3.2}
As shown in Eq.~(\ref{eq:112}) and Eq.~(\ref{eq:244}), while the second-order \acp{GW} $h^{(2)}_{lm}$ and the fourth-order \acp{GW} $h^{(4)}_{lm}$ have identical equations of motion, their explicit expressions vary due to the differences in the first-order scalar perturbation $\phi^{(1)}$ and the second-order scalar perturbation $\phi^{(2)}$ in the source terms. The vertex rules associated with the vertex diagram in Fig.~\ref{fig:1} describe the relationship between $n$-th order induced \acp{GW} and lower-order perturbations. However, it does not directly provide the explicit expression for $n$-th order induced \acp{GW}. A single vertex diagram is insufficient to fully describe higher-order induced \acp{GW}.

\begin{figure}[htbp]
    \centering
\captionsetup{justification=centering,singlelinecheck=false}
\hspace{0.1\columnwidth}   
       \includegraphics[width=0.95\columnwidth]{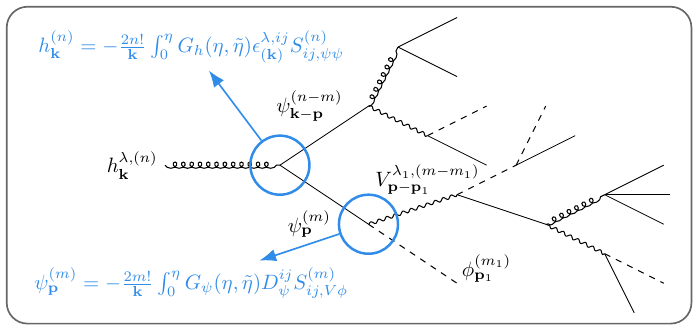}
        \hspace{0.1\columnwidth}        
\caption{\label{fig:Coral} The tree diagram of induced \acp{GW}. $G_{h}\left(\eta,\bar{\eta}  \right)$ and $G_{\psi}\left(\eta,\bar{\eta}  \right)$ represent the Green's functions of tensor perturbation $h_{ij}$ and scalar perturbation $\psi$, respectively.}
\end{figure}

In conventional approaches, obtaining the explicit expression for n-th order induced \acp{GW} involves solving the \ac{CP} equations iteratively, starting from first-order perturbations. In the vertex approach, we allow the branches on the right side of the vertex diagram to bifurcate into new vertices, resembling the growth of a tree, until all emerging branches correspond exclusively to first-order perturbations. As illustrated in Fig.~\ref{fig:Coral}, since the analytical solutions of the first-order perturbations are known, the ‘fully grown’ tree diagram allows us to systematically derive the expression for higher-order induced \acp{GW} using the vertex approach. By summing up these tree diagrams, we obtain the explicit expression for the $n$-th order induced \acp{GW}. 

\section{Loop diagrams}\label{sec:4.0}
Calculating the two-point correlation function of \acp{GW} yields their energy density spectrum, a key observable in \ac{SGWB} experiments. Within the vertex approach, the explicit expression for induced \acp{GW} can be directly derived from tree diagrams.  The two-point correlation function of induced \acp{GW} corresponds to loop diagrams formed by connecting two tree diagrams. Different connection patterns between these two tree diagrams lead to distinct types of loop diagrams, which align with the Wick expansion of the multi-point correlation function of primordial perturbations.

Previous studies have systematically analyzed second-order \acp{SIGW}, which correspond to the one-loop calculations of \acp{CP} \cite{Domenech:2021ztg}. As highlighted earlier, due to the large amplitude of primordial perturbations on small scales, higher-order effects are significant in small-scale cosmological observations. Refs.~\cite{Yuan:2019udt,Chen:2019xse,Chang:2022nzu,Chang:2023vjk} show that third-order \acp{SIGW} greatly influence the total energy density spectrum of \acp{SIGW}, both for narrow-band monochromatic and broad-band log-normal primordial power spectra. For the log-normal spectrum, third-order corrections enhance the signal-to-noise ratio of \ac{LISA} by approximately $70\%$ and reduce the upper limit on the amplitude $A_{\zeta}$ from \ac{PTA} observations by around $50\%$ \cite{Chang:2023vjk}. Therefore, to accurately determine cosmological parameters on small scales, it is imperative to systematically calculate or rigorously estimate the contributions of higher-order \acp{CP}. For \acp{SIGW}, the total energy density spectrum can be expressed in the following form
\begin{equation}\label{eq:WRZK}
\begin{aligned}
\Omega_{\mathrm{GW}}(k)&=A_\zeta^2\left(\Omega_{\mathrm{GW}}^{(2,2)}(k)\right)\bigg|_{A_\zeta=1}\\
&+A_\zeta^3\left(\sum^{1}_{m=0}~\Omega_{\mathrm{GW}}^{(3+m,3-m)}(k)\right)\bigg|_{A_\zeta=1} \\
&+A_\zeta^4\left(\sum^{2}_{m=0}~\Omega_{\mathrm{GW}}^{(4+m,4-m)}(k)\right)\bigg|_{A_\zeta=1}\\
&+\sum^{+\infty}_{n=5} A_\zeta^n\left(\sum^{n-2}_{m=0}~\Omega_{\mathrm{GW}}^{(n+m,n-m)}(k)\right)\bigg|_{A_\zeta=1}  \ , 
\end{aligned}
\end{equation}
where, $\left(\Omega_{\mathrm{GW}}^{(n,m)}(k)\right)\bigg|_{A_\zeta=1}$ denotes the energy density spectrum corresponds to the two-point correlation function $\langle h^{\lambda,(n)}_{\mathbf{k}} h^{\lambda ',(m)}_{\mathbf{k}'}\rangle$ when the amplitude of the primordial power spectrum $A_\zeta = 1$. It should be noted that the contribution of the $n$-th order energy density spectrum of \acp{SIGW} does not solely come from the $n$-th order \acp{SIGW}. Taking the fourth-order \acp{SIGW} as an example, both the two-point correlation functions $\langle h^{\lambda,(3)}_{\mathbf{k}} h^{\lambda ',(3)}_{\mathbf{k}'}\rangle$ and $\langle h^{\lambda,(4)}_{\mathbf{k}} h^{\lambda ',(2)}_{\mathbf{k}'}\rangle$ affect the fourth-order energy density spectrum. This is the reason for the summation over $m$ in Eq.~(\ref{eq:WRZK}).

In this section, we will perform a general analysis of the loop diagram structure of induced \acp{GW} and investigate the differences between \acp{SIGW} and \acp{TSIGW} at the \ac{1PR} diagrams level.

\subsection{General structure of loop diagrams}\label{sec:4.1}
In Sec.~\ref{sec:3.0}, we introduced the vertex rules for calculating higher-order induced \acp{GW}. This enables us to derive the explicit expressions for higher-order induced \acp{GW} directly. For instance, the $n$-th order \acp{SIGW} can be formally represented as
\begin{equation}\label{eq:nhf}
\begin{aligned}
h^{\lambda,(n)}_{\mathbf{k}}&(\eta)=\sum_{i=1}^{n_i}h_{i,\mathbf{k}}^{\lambda,(n)}(\eta)\\
&= \sum_{i=1}^{n_i} \int \frac{\mathrm{d}^3 p_1}{(2 \pi)^{3 / 2}} \cdots \int \frac{\mathrm{d}^3 p_{n-1}}{(2 \pi)^{3 / 2}}  \\
&\times \varepsilon^{\lambda, l m}(\mathbf{k})\mathbb{P}^{i}_{lm}\left( \mathbf{k}, \mathbf{p}_1, \mathbf{p}_2, \cdots, \mathbf{p}_{n-1} \right) \\
&\times I_{i}^{(n)}\zeta_{\mathbf{k}-\mathbf{p}_1} \zeta_{\mathbf{p}_1-\mathbf{p}_2}\cdots \zeta_{\mathbf{p}_{n-1}} \ ,
\end{aligned}
\end{equation}
where the momentum polynomials $\mathbb{P}^{i}_{lm}$ and the $n$-th order kernel functions $ I_{i}^{(n)}$ are determined by the specific configurations of the tree diagrams. The summation symbol in Eq.~(\ref{eq:nhf}) denotes the summation over all possible tree diagram structures.

To obtain the energy density spectrum of the $n$-th order \acp{SIGW}, we need to calculate the two-point correlation function of the \acp{SIGW}, which corresponds to the calculation of the loop diagrams formed by connecting the tree diagrams. The two-point correlation function of $n$-th order \acp{SIGW} in Eq.~(\ref{eq:nhf}) can be expressed as follows
\begin{equation}\label{eq:dis}
\begin{aligned}
&\langle h^{\lambda,(n)}_{\mathbf{k}} h^{\lambda',(n)}_{\mathbf{k}'} \rangle=\sum_{i=1}^{n_i}\sum_{i'=1}^{n'_i} \int \frac{\mathrm{d}^3 p_1}{(2 \pi)^{3 / 2}} \cdots \int \frac{\mathrm{d}^3 p_{n-1}}{(2 \pi)^{3 / 2}}  \\
&\times\int \frac{\mathrm{d}^3 p'_1}{(2 \pi)^{3 / 2}} \cdots \int \frac{\mathrm{d}^3 p'_{n-1}}{(2 \pi)^{3 / 2}}\varepsilon^{\lambda, l m}(\mathbf{k})\varepsilon^{\lambda', rs}(\mathbf{k}')\\
&\times\mathbb{P}^{i}_{lm}\left( \mathbf{k}, \cdots, \mathbf{p}_{n-1} \right)\mathbb{P}^{i'}_{rs} \left( \mathbf{k}', \cdots, \mathbf{p}'_{n-1} \right) I^{(n)}_i I^{(n)}_{i'} \\
&\times \langle \zeta_{\mathbf{k}-\mathbf{p}_1} \cdots \zeta_{\mathbf{p}_{n-1}} \zeta_{\mathbf{k}'-\mathbf{p}'_1} \cdots \zeta_{\mathbf{p}'_{n-1}}   \rangle \ ,
\end{aligned}
\end{equation}
where the physical quantities with and without primes respectively correspond to the two \acp{SIGW} in the correlation function. To better demonstrate the general loop diagram structure of \acp{SIGW}, we break down the expression in Eq.~(\ref{eq:dis}) into the following three components:

\noindent
\textbf{a, The integrand:} containing the momentum polynomial $\varepsilon^{\lambda, l m}(\mathbf{k})\varepsilon^{\lambda', rs}(\mathbf{k}')\mathbb{P}^{i}_{lm}\mathbb{P}^{i'}_{rs}$ and the product of two kernel functions $I^{(n)}_i I^{(n)}_{i'}$. The explicit expression for the integrand can be directly derived using vertex rules and tree diagrams.

\noindent
\textbf{b, $2n$-point correlation function:} in the calculation of the $n$-th order energy density spectrum, the $2n$-point correlation function of the primordial curvature perturbations appears. For Gaussian primordial perturbations, Wick's theorem allows us to expand it into the product of various two-point correlation functions of the primordial curvature perturbations. Different Wick's theorem expansions lead to different forms of loop diagrams. Additionally, we can use the definition of the primordial power spectrum: $\left\langle\zeta_{\mathbf{k}} \zeta_{\mathbf{k}^{\prime}}\right\rangle=\frac{2 \pi^2}{k^3} \delta\left(\mathbf{k}+\mathbf{k}^{\prime}\right) \mathcal{P}_\zeta(k)$, to express the $2n$-point correlation function as the product of $n$ primordial power spectra.

\noindent
\textbf{c, Momentum integrals:} the calculation of the $n$-th order energy density spectrum of \acp{SIGW} directly involves $2(n-1)$ three-dimensional momentum integrals. Using the $n-1$ three-dimensional $\delta$ functions in the definition of the primordial power spectrum, we can eliminate $n-1$ integrals (the remaining three-dimensional $\delta$ function $\delta\left(\mathbf{k}+\mathbf{k}' \right)$ represents the momentum conservation of the \acp{SIGW}).

By analyzing the structure of loop diagrams for \acp{SIGW}, we discover that the explicit expressions for the momentum polynomials and kernel functions in the integrand can be obtained directly using vertex rules and tree diagrams. The complexity of loop diagram calculations mainly arises from two aspects: 

\noindent
\textbf{1,} The expansion of the Wick theorem produces numerous different loop diagram structures, resulting in an excessive number of loop diagrams to be calculated. 

\noindent
\textbf{2,} Choosing specific coordinate systems and appropriate integration variables to calculate the $n-1$ three-dimensional momentum integrals.

As an important example in the study of loop diagram structures, we will examine the \ac{1PR} diagrams for induced \acp{GW} in the next subsection. We will see that the \ac{1PR} diagrams do not affect \acp{SIGW}.

\subsection{\acp{1PR} diagrams of induced \acp{GW}}\label{sec:4.2}
Similar to the large-amplitude primordial curvature perturbations on small scales, primordial tensor perturbations on small scales may also exhibit large amplitudes. The higher-order \acp{GW} induced by the combined effects of small-scale primordial tensor and scalar perturbations are referred to as \acp{TSIGW} \cite{Chang:2022vlv,Wu:2024qdb,Yu:2023lmo,Bari:2023rcw,Picard:2023sbz}. The presence of primordial tensor perturbations introduces notable differences in the loop structures of \acp{TSIGW} compared to \acp{SIGW}. To study the differences between \acp{SIGW} and \acp{TSIGW} in loop diagram calculations, we concentrate on the simplest loop diagram structure: \ac{1PR} diagrams in third-order induced \acp{GW}. In the calculation of two-loop diagrams of \acp{SIGW}, \ac{1PR} diagrams have no impact on the two-point correlation function. 
\begin{figure}[htbp]
    \centering
    \captionsetup{justification=centering,singlelinecheck=false}
\hspace{0.1\columnwidth} 
    \subfloat[]{\includegraphics[width=1\columnwidth]{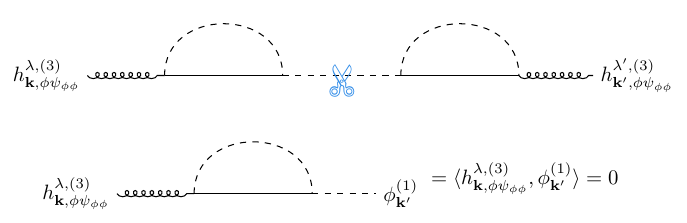}\label{fig:hphi} } \\
    \hspace{0.1\columnwidth}
   \subfloat[]{\includegraphics[width=1\columnwidth]{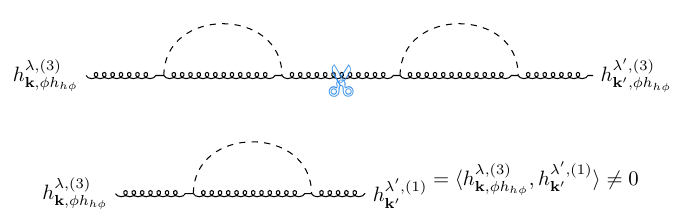}\label{fig:hh} }
\caption{ The \ac{1PR} diagrams resulting from the contraction of two-branch tree diagrams of third-order induced \acp{GW}.  }
\end{figure}
As shown in Fig.~\ref{fig:hphi}, the two-point correlation function of third-order \acp{SIGW} associated with the \ac{1PR} diagram can be expressed as
\begin{equation}\label{eq:h3h3}
	\begin{aligned}
		 \delta&\left(\mathbf{k}+\mathbf{k}'  \right) \frac{2\pi^2}{k^3}  \langle h^{\lambda,(3)}_{\mathbf{k},\phi\psi_{\phi\phi}} h^{\lambda',(3)}_{\mathbf{k}',\phi\psi_{\phi\phi}}\rangle \\
  &=\langle h^{\lambda,(3)}_{\mathbf{k},\phi\psi_{\phi\phi}}\phi^{(1)}_{\mathbf{k}'} \rangle  ~\frac{1}{\mathcal{P}_{\zeta}(k)T^{2}_{\phi}\left( k'\eta \right) }~ \langle  \phi^{(1)}_{\mathbf{k}} h^{\lambda',(3)}_{\mathbf{k}',\phi\psi_{\phi\phi}}\rangle \ ,
	\end{aligned}
\end{equation}
where
\begin{align}
	\langle h^{\lambda,(3)}_{\mathbf{k},\phi\psi_{\phi\phi}}\phi^{(1)}_{\mathbf{k}'} \rangle &=\int\frac{\mathrm{d}^3p}{(2\pi)^{3/2}}\int\frac{\mathrm{d}^3q}{(2\pi)^{3/2}}\varepsilon^{\lambda,lm}(\mathbf{k})p_lp_m \nonumber\\
	&\frac{16}{81} I^{(3)}_{\phi\psi_{\phi\phi}}(|\mathbf{k}-\mathbf{p}|,|\mathbf{p}-\mathbf{q}|,|\mathbf{q}|,\eta)  T_{\phi}(k' \eta) \nonumber\\
    &\langle \zeta_{\mathbf{k}-\mathbf{p}} \zeta_{\mathbf{p}-\mathbf{q}} \rangle\langle\zeta_{\mathbf{q}}\zeta_{\mathbf{k}'} \rangle \ .
	\label{eq:H3} 
\end{align}
In Eq.~(\ref{eq:h3h3}) and Eq.~(\ref{eq:H3}), $\zeta_{\mathbf{k}}$ represents the primordial curvature perturbation, and $\mathcal{P}_{\zeta}(k)$ is the corresponding primordial power spectrum. $I^{(3)}_{\phi\psi_{\phi\phi}}$ is the third-order kernel function \cite{Zhou:2021vcw}.  $T_{\phi}(k\eta)$ is the transfer function of first-order scalar perturbation \cite{Inomata:2020cck}. Here, the \ac{1PR} diagram of third-order \acp{SIGW}  is decomposed into the product of two \ac{1PI} diagrams. Therefore, we only need to consider the two-point function $\langle   h^{\lambda,(3)}_{\mathbf{k}',\phi\psi_{\phi\phi}} \phi^{(1)}_{\mathbf{k}}\rangle$. In Eq.~(\ref{eq:H3}), the momentum integral can be calculated in a spherical coordinate system with $\mathbf{k}$ as the $z$-axis.  In this case, the contraction of the polarization tensor $\varepsilon^{\lambda,lm}(\mathbf{k})$ with momenta $p_l$ and $p_m$ can be expressed as \cite{Domenech:2021ztg}
\begin{equation}
\varepsilon^{\lambda,lm}(\mathbf{k})p_lp_m=\frac{p^2}{\sqrt{2}} \sin ^2 \theta_p \times \begin{cases}\cos 2 \phi_p, & \lambda=+ \\ \sin 2 \phi_p, & \lambda=\times\end{cases} \ .
\end{equation}
Notably, the kernel function $I^{(3)}_{\phi\psi_{\phi\phi}}$ in Eq.~(\ref{eq:H3}) relies solely on the relative positions of $\mathbf{k}$, $\mathbf{p}$, and $\mathbf{q}$, completely independent of the overall azimuth angle $\phi_p$. Integrating over  $\phi_p$ yields $\int_0^{2\pi} \varepsilon^{\lambda,lm}(\mathbf{k})p_lp_m \mathrm{d}\phi_p =0 $. Consequently,  the two-point correlation function $\langle   h^{\lambda,(3)}_{\mathbf{k}',\phi\psi_{\phi\phi}} \phi^{(1)}_{\mathbf{k}}\rangle=0$, and the \ac{1PR} diagrams of third-order \acp{SIGW} do not contribute to the energy density spectrum. This conclusion can be generalized to arbitrary higher-order \acp{SIGW} with different kinds of source terms. However, the above results are not applicable to \acp{TSIGW}.  For example, as shown in Fig.~\ref{fig:hh}, the \ac{1PR} diagram for third-order \acp{TSIGW} can be decomposed into the product of two \ac{1PI} diagrams. In this case, the polarized tensor $\varepsilon^{\lambda,lm}(\mathbf{k})$ in the two-point correlation function $\langle h^{\lambda,(3)}_{\mathbf{k}}h^{\lambda_1,(1)}_{\mathbf{k}'} \rangle$ will contract with other polarized tensors \cite{Chen:2022dah}. The corresponding two-point function can be written as
\begin{equation}\label{eq:1PRh}
	\begin{aligned}
		 \delta&\left(\mathbf{k}+\mathbf{k}'  \right) \frac{2\pi^2}{k^3} \langle h^{\lambda,(3)}_{\mathbf{k},\phi h_{\phi h}} ~h^{\lambda',(3)}_{\mathbf{k}',\phi h_{\phi h}}\rangle\\
  &=\langle  h^{\lambda,(3)}_{\mathbf{k},\phi h_{\phi h}} h^{\lambda_1,(1)}_{\mathbf{k}'}\rangle \frac{\delta_{\lambda_{1} \lambda'_{1}}}{\mathcal{P}_{h}(k')T^{2}_{h}\left( k'\eta \right) }~ \langle  h^{\lambda'_{1},(1)}_{\mathbf{k}} h^{\lambda',(3)}_{\mathbf{k}',\phi h_{\phi h}}\rangle \ ,
	\end{aligned}
\end{equation}
where $\mathcal{P}_{h}(k')$ and $T^{2}_{h}\left( k'\eta\right)$ represent the primordial power spectrum and the first-order transfer function of tensor perturbation, respectively. The explicit expressions of $h^{\lambda,(3)}_{\mathbf{k}, s h_{s h}}$$\left(s=\phi,~ \psi  \right)$ are given in Appendix.~\ref{sec:C}. As shown in Fig.~\ref{fig:all}, in the case of the log-normal primordial power spectra: $A_{\zeta}\left(2\pi\sigma_\zeta^2\right)^{-1/2}\exp{(-\ln^2{(k/k_*)}/(2\sigma_\zeta^2))}$ and $A_{h}\left(2\pi\sigma_h^2\right)^{-1/2}\exp{(-\ln^2{(k/k_*)}/(2\sigma_h^2))}$, the \ac{1PR} diagram of $h^{\lambda,(3)}_{\mathbf{k}, s h_{s h}}$  will have a non-zero impact on the total energy density spectrum of induced \acp{GW}.
\begin{figure}[htbp]
    \centering
    \includegraphics[width=0.95\columnwidth]{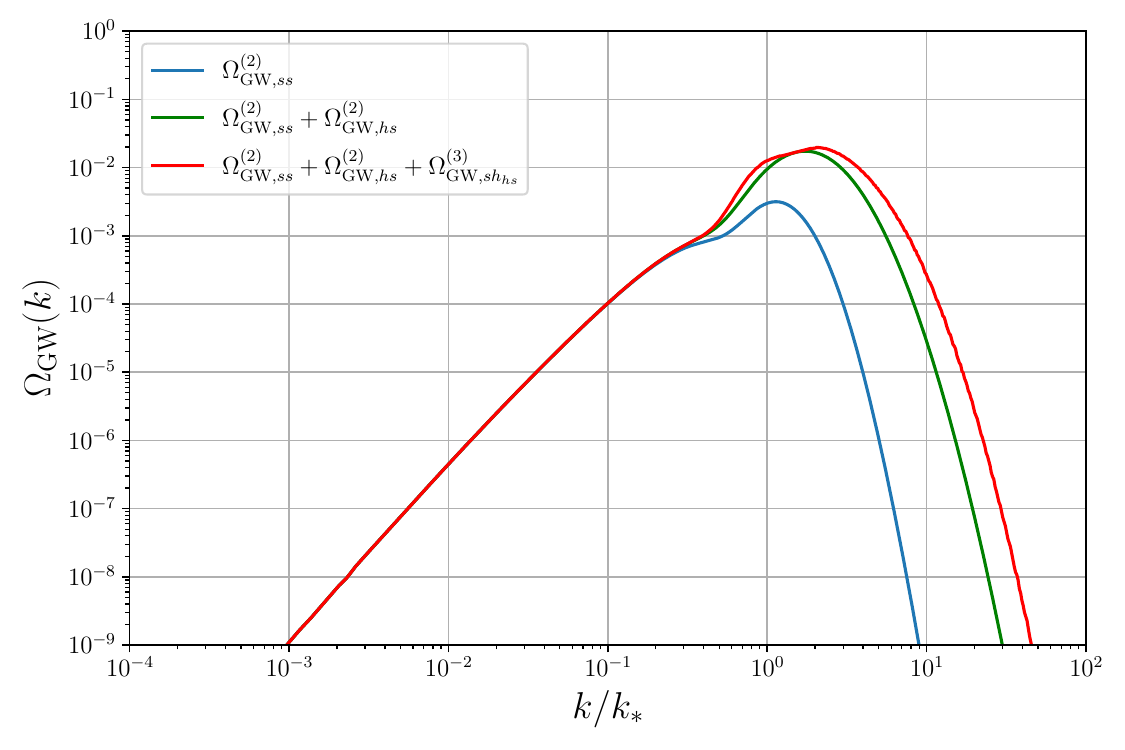}
\caption{\label{fig:all} The blue line, green line, and red line correspond to the energy density spectra of $h^{\lambda,(2)}_{\mathbf{k},ss}$, $h^{\lambda,(2)}_{\mathbf{k},ss}$+$h^{\lambda,(2)}_{\mathbf{k},sh}$, and $h^{\lambda,(2)}_{\mathbf{k},ss}$+$h^{\lambda,(2)}_{\mathbf{k},sh}$+\ac{1PR} diagrams of $h^{\lambda,(3)}_{\mathbf{k},sh_{sh}}$, respectively. The symbols $s=\phi,\psi$ and $h$ represent scalar perturbations and tensor perturbation, respectively. We set $A_{\zeta} = 0.1$, $A_h=0.05$ and $\sigma_\zeta = \sigma_h = 0.5$.} 
\end{figure}

\begin{figure}[htbp]
    \centering
    \includegraphics[width=0.99\columnwidth]{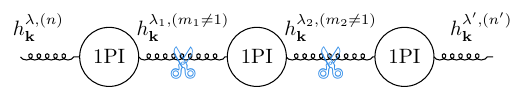}
\caption{\label{fig:ineq} In higher-order \ac{CP} theory, the decomposition of \ac{1PR} diagrams requires consideration of the order of perturbations. }
\end{figure}

It should be noted that in higher-order \ac{CP} theory, the curves in tree and loop diagrams also encode the order of \acp{CP}. When decomposing a \ac{1PR} diagram into multiple \ac{1PI}  diagrams, it is crucial to consider the order of the perturbations associated with the 'cut' curves. As shown in Fig.~\ref{fig:ineq}, the \acp{1PR} diagram cannot be simply expressed as $ \langle h^{\lambda,(n)}_{\mathbf{k}} h^{\lambda',(n')}_{\mathbf{k}'}\rangle \sim  \langle h^{\lambda,(n)}_{\mathbf{k}} h^{\lambda',(1)}_{\mathbf{k}'}\rangle^3$. The correct decomposition obtained through direct calculation is
\begin{equation}
	\begin{aligned}
		 \langle h^{\lambda,(n)}_{\mathbf{k}} h^{\lambda',(n')}_{\mathbf{k}'}\rangle ~ \sim ~& \langle h^{\lambda,(n)}_{\mathbf{k}} h^{\lambda_1,(m_1)}_{\mathbf{k}'}\rangle \langle h^{\lambda_1,(m_1)}_{\mathbf{k}} h^{\lambda_2,(m_2)}_{\mathbf{k}'}\rangle \\
   &\langle h^{\lambda_2,(m_2)}_{\mathbf{k}} h^{\lambda',(n')}_{\mathbf{k}'}\rangle \ .
	\end{aligned}
\end{equation}
As shown in Table.~\ref{ta:1}, the physical implications of tree diagrams and loop diagrams in classical higher-order \acp{CP} are different from those of Feynman diagrams in \ac{QFT}. All the calculation techniques for tree diagrams and loop diagrams involved here must be derived or proven from basic definitions.
\begin{table}[h!]
\centering
\begin{tabular}{|p{2.4cm}|p{2.4cm}|p{2.3cm}|}
\hline
& ~ \ac{QFT} &~ Classical \acp{CP}   \\
\hline
 Origin  & ~Dyson series &~\ac{CP} equations \\
\hline
Tree diagram &Lowest-order contribution & Explicit expression of \acp{CP} \\
\hline
Loop diagram & Higher-order 
correction &  Two-point function \\
\hline
External lines&~ Particles &  ~ \acp{CP}  \\
\hline
\end{tabular}
\caption{The distinction between Feynman diagrams in \ac{QFT} and diagrams in classical \ac{CP} theory. }
\label{ta:1}
\end{table}

\section{Conclusion and discussion}\label{sec:5.0}
Over the past few years, considerable interest has focused on large primordial perturbations at small scales. The large primordial perturbations play a pivotal role in areas such as \acp{PBH} and induced \acp{GW}. In this scenario, the influence of higher-order \acp{CP} on small-scale cosmological observations is non-negligible. If higher-order \acp{CP} are not systematically calculated or rigorously estimated, and only lower-order perturbations are considered, the cosmological parameters derived from current observations, such as the small-scale primordial power spectrum, will significantly deviate from their true values. Systematically investigating the effects of higher-order \acp{CP} during various dominant epochs and in different gauges on small-scale cosmological observations is undoubtedly a crucial and fundamental challenge. 

Traditional \ac{CP} theory requires order-by-order derivation and solution of higher-order perturbation equations to obtain the formal solutions of higher-order \acp{CP}. The complexity and tediousness of these calculations make it difficult to apply conventional methods to the study of arbitrary higher-order \acp{CP}.  We proposed the vertex approach for directly constructing higher-order \acp{CP} from the equations of motion of lower-order \acp{CP}. Utilizing tree diagrams and vertex rules, one can derive the specific expressions for higher-order \acp{CP}. In this study, we focus on the higher-order induced \acp{GW} during the \ac{RD} era in Newtonian gauge, the theoretical framework of vertex approach can be extended to various types of \acp{CP}, different dominant epochs of the universe, and different gauges.

In investigating induced \acp{GW}, solving higher-order \acp{CP} is only part of the task. We also need to compute two-point correlation functions to derive the energy density spectrum, represented by loop diagrams connecting different tree diagrams. In this paper, we studied the general structure of loop diagrams. The results indicate that the complexity of loop diagram calculations primarily arises from the Wick theorem expansion and the computation of momentum integrals. Unlike the complexity in higher-order \ac{CP} equations caused by redundancy in traditional methods, the complexity of loop diagram calculations involves the specific computation process of three-dimensional momentum integrals in particular coordinate systems. Similar to the calculation of loop diagrams in quantum field theory, we need to develop more systematic methods for loop diagram calculations in \ac{CP} theory to improve the calculation efficiency of higher-order induced \acp{GW}. Efficiently calculating momentum integrals in \ac{CP} theory requires further investigation.

\vspace{0.3cm}
\begin{acknowledgements} 
The authors want to thank Dr. Fei-Yu Chen and Dr. Quan-feng Wu for useful discussions. This work has been funded by the National Nature Science Foundation of China (NSFC) under grant No. 12447127, No. 12075249, No. 11690022, No. 12475075, No. 11935009, and No. 12375052, and the Key Research Program of the Chinese Academy of Sciences under Grant No. XDPB15.
\end{acknowledgements}

\onecolumngrid
\appendix
\section{Higher order cosmological perturbations }\label{sec:A}

To determine the source terms associated with various vertex diagrams, We need to calculate the $n$-th order perturbations of Einstein's field equations without performing a perturbative expansion of the \acp{CP}. The perturbation of the metric $g_{\mu\nu}$ in \ac{FLRW} background is given by
\begin{equation}\label{eq:gx}
    \begin{aligned}
    g^{(0)}_{\mu\nu}&=a^2\eta_{\mu\nu}    \ , \\
      \delta g_{\mu \nu} &=a^2h_{\mu\nu}-a^2V_{\nu}n_{\mu}-a^2V_{\mu}n_{\nu}-2a^2n_{\mu}n_{\nu}\phi-2a^2\psi m_{\mu\nu} \ , \\
      &=a^2
\left(
  \begin{array}{ccc}
    -2\phi  & \partial_iB +V_i  \\
    \partial_iB +V_i  & -2\psi \delta_{ij}+h_{ij} \\
  \end{array}
\right) \ , \\
      \delta^{n} g_{\mu \nu}&=0 \ , \  \left(n=2,3,4,\cdots \right) \ ,
    \end{aligned}
\end{equation}
where the symbol $\delta^n g_{\mu\nu}=g_{\mu\nu}^{(n)}$ represents the $n$-th order perturbation of $g_{\mu\nu}$. $\phi$ and $\psi$ are first order scalar perturbations. $h_{ij}$ and $V_i$ represent tensor perturbations and vector perturbations, respectively. Here, we have defined
\begin{equation}
h_{\mu\nu} =
\left(
  \begin{array}{ccc}
    0  & 0  \\
    0  & h_{ij} \\
  \end{array}
\right) \  ,  \ m_{\mu\nu} =
\left(
  \begin{array}{ccc}
    0  & 0  \\
    0  & \delta_{ij} \\
  \end{array}
\right) \  ,  \  V_{\mu}=\left(0, V_{i}  \right)         \ , \     n_{\mu} = \left(-1,0,0,0 \right)  \  ,  \  n^{\mu} = \left(1,0,0,0 \right) \ .
\end{equation}
To derive the vertex rules associated with multi-branch vertices, it is necessary to use the metric perturbations in Eq.~(\ref{eq:gx}) to calculate the $n$-th order perturbation of the Einstein tensor. The Christoffel symbol can be expressed as
\begin{equation}
    \begin{aligned}
        \Gamma^{\sigma}_{\mu\nu}=\frac{1}{2}g^{\sigma \alpha}\left(\partial_{\nu}g_{\alpha \mu}+\partial_{\mu}g_{\alpha \nu}-\partial_{\alpha}g_{\mu \nu}  \right) \ .
    \end{aligned}
\end{equation}
The $n$-th order perturbation of the Christoffel symbol can be represented using the perturbations of the metrics $g_{\mu\nu}$ and $g^{\mu\nu}$
\begin{equation}\label{eq:Gamma}
    \begin{aligned}
        \delta^n \Gamma^{\sigma}_{\mu\nu}&=\frac{1}{2} \left(\delta^{n-1} g^{\sigma \alpha}\right)\delta\left(\partial_{\nu}g_{\alpha \mu}+\partial_{\mu}g_{\alpha \nu}-\partial_{\alpha}g_{\mu \nu}  \right)+\frac{1}{2}\left(\delta^n g^{\sigma \alpha}\right)\left(\partial_{\nu}g^{(0)}_{\alpha \mu}+\partial_{\mu}g^{(0)}_{\alpha \nu}-\partial_{\alpha}g^{(0)}_{\mu \nu}  \right) \\
        &=\frac{1}{2} \left(\delta^{n-1} g^{\sigma \alpha}\right)\left[\partial_{\nu}\left(a^2h_{\alpha\mu}-a^2V_{\mu}n_{\alpha}-a^2V_{\alpha}n_{\mu}-2a^2n_{\alpha}n_{\mu}\phi-2a^2\psi m_{\alpha\mu}  \right)+\partial_{\mu}\left(a^2h_{\alpha\nu}-a^2V_{\nu}n_{\alpha} \right.\right. \\
        &\left.\left.-a^2V_{\alpha}n_{\nu}-2a^2n_{\alpha}n_{\nu}\phi-2a^2\psi m_{\alpha\nu}   \right)
        -\partial_{\alpha}\left(a^2h_{\mu\nu}-a^2V_{\nu}n_{\mu}-a^2V_{\mu}n_{\nu}-2a^2n_{\mu}n_{\nu}\phi-2a^2\psi m_{\mu\nu}  \right)  \right] \\
        &+aa'\left(\delta^n g^{\sigma \alpha}\right)\left(-\eta_{\alpha \mu}n_{\nu}-\eta_{\alpha \nu}n_{\mu}+\eta_{\mu\nu}n_{\alpha} \right) \ .
    \end{aligned}
\end{equation}
In Eq.~(\ref{eq:Gamma}), the metric perturbation $\delta^n g^{\sigma\alpha}$ can be straightforwardly determined from the relation $g^{\sigma\alpha}g_{\alpha\beta}=\delta^{\sigma}_{\beta}$. The explicit expression of $\delta^n g^{\sigma\alpha}$ is given by
\begin{equation}
    \delta^n g^{\sigma\alpha}=\delta^n g^{\sigma\alpha}_{ss}+\delta^n g^{\sigma\alpha}_{hh}+\delta^n g^{\sigma\alpha}_{sh}+\delta^n g^{\sigma\alpha}_{vv}+\delta^n g^{\sigma\alpha}_{sv}+\delta^n g^{\sigma\alpha}_{vh} \ ,
\end{equation}
where
\begin{equation}\label{eq:gss}
    \begin{aligned}
        \delta^n g^{\sigma\alpha}_{ss}=\frac{2^n}{a^2}\left(\left( 
-1 \right)^{n+1}\phi^nn^{\sigma}n^{\alpha}+\psi^n m^{\sigma\alpha} \right) \ ,
    \end{aligned}
\end{equation}
\begin{equation}\label{eq:ghh}
    \begin{aligned}
        \delta^n g^{\sigma\alpha}_{hh}=\frac{(-1)^n}{a^2}\begin{cases}~  h^{\alpha}_{i} h^{\sigma i}  \  , & n=2 \ , \\
        \\
        ~ h_{bc}h^{\sigma b} h^{\alpha c}  \  ,  & n=3 \ , \\
        \\
        ~h^{i_1}_{b} h_{i_1 c} h^{\sigma b } h^{\alpha c }\ , &  n=4  \ ,\\
        \\
         ~h^{i_1}_{b}h^{i_2}_{c} h_{i_1 i_2} h^{\sigma b } h^{\alpha c }\ , &  n=5  \ ,\\
        \\
        ~h^{i_1}_{b} h^{i_2}_{c}  h_{i_1}^{i_3} h_{i_2}^{i_4} \cdots h_{i_{n-5}i_{n-4}}  h^{\sigma b } h^{\alpha c }\ , &  n>5  \ , \end{cases}
    \end{aligned}
\end{equation}
\begin{equation}\label{eq:gsh}
    \begin{aligned}
        \delta^n g^{\sigma\alpha}_{sh}=\begin{cases} ~0 \ , & n=1 \ , \\
        \\
        ~\sum^{n-1}_{m=1}\left( -1 \right)^m 2^m C^m_n\psi^m \left( \delta^{n-m} g^{\sigma\alpha}_{hh}\right)\ , &  n>1   \ , \end{cases}
    \end{aligned}
\end{equation}
\begin{equation}\label{eq:gvv}
    \begin{aligned}
        \delta^n g^{\sigma\alpha}_{vv}=\begin{cases} ~\frac{\left( -1 \right)^{\frac{n-1}{2}}}{a^2} \left(V^i V_i\right)^{\frac{n-1}{2}}\left(V^{\sigma}n^{\alpha} +V^{\alpha}n^{\sigma}\right) \ , & n=odd \ , \\
        \\
        ~\frac{1}{a^2}\left( \left( -1 \right)^{n/2}\left( V^i V_i \right)^{n/2-1}V^{\alpha}V^{\sigma}+\left( -1 \right)^{n/2+1}\left( V^i V_i \right)^{n/2}n^{\alpha}n^{\sigma}  \right)\ , &  n=even   \ , \end{cases}
    \end{aligned}
\end{equation}
\begin{equation}\label{eq:gsv}
    \begin{aligned}
        \delta^n g^{\sigma\alpha}_{sv}=\frac{1}{a^2}\begin{cases} 
        ~ 0 \ , &  n=1  \ , \\
        \\
        ~ 2\left(V^{\sigma}n^{\alpha}+2V^{\alpha}n^{\sigma}\right)\left( \psi-\phi \right)\ , & n=2 \ , \\
        \\
        ~ 4\left(V^{\alpha}n^{\sigma}+V^{\sigma}n^{\alpha}\right)\left(\phi^2+\psi^2-\phi\psi  \right)+2V^{\alpha}V^{\sigma}\left(\phi-2\psi  \right)\\
        ~+2V^aV_a n^{\sigma}n^{\alpha}\left(\psi-2\phi  \right)   \ ,  &  n=3   \ , \\
        \\
         ~4  V_{b}\ V^{b}\left(V^\alpha n^\sigma\left(  \phi-  \psi\right)+n^\alpha\left(V^\sigma\left(  \phi-  \psi\right)+n^\sigma\left(3\phi^2-2\phi\psi+\psi^2\right)\right)\right) \\
         ~+8V^\alpha n^\sigma\left(-\phi^3+\phi^2\psi-\phi\psi^2+\psi^3\right)-4V^\sigma\left(V^\alpha\left(\phi^2-2\phi\psi+3\psi^2\right)\right.\\
         ~\left.+2 n^\alpha\left( \phi^3-\phi^2\psi+\phi\psi^2-\psi^3\right)\right) \ ,  &  n=4   \ , \\
        \\
        ~~\cdots \cdots\cdots  \ , 
        \end{cases}
    \end{aligned}
\end{equation}
\begin{equation}\label{eq:gvh}
    \begin{aligned}
        \delta^n g^{\sigma\alpha}_{vh}=\frac{1}{a^2}\begin{cases} 
        ~ 0 \ , &  n=1  \ , \\
        \\
        ~ -V^b\left(h^{\alpha}_bn^{\sigma}+h^{\sigma}_b n^{\alpha}   \right)\ , & n=2 \ , \\
        \\
        ~ V^b\left( V^{\alpha}h^{\sigma}_b+V^{\sigma}h^{\alpha}_b+h_{bc}h^{\sigma c}n^{\alpha}+h_{bc}h^{\alpha c}n^{\sigma}  \right)-V^bV^ch_{bc}n^{\sigma}n^{\alpha}  \ ,  &  n=3   \ , \\
        \\
        ~ V^b\left(-  V^\sigma  h_{b c} h^{\alpha  c}+ V^{\alpha }  h_{b c}\left(- h^{\sigma c}+ V^c n^{\sigma}\right)-h_b^d h_{c d}\left(  h^{\alpha c} n^{\sigma}+  h^{\sigma c} n^{\alpha}\right)\right. \\
        ~ \left.  +  V^c\left(-  h^{\sigma}_b  h_c^\alpha+\left( V^\sigma\  h_{b c}+ h_{b}^d h_{c d} n^{\sigma}\right) n^{\alpha}+  V_b\left( h_c^\alpha n^{\sigma}+  h^\sigma_c n^{\alpha}\right)\right)\right)\ ,  &  n=4  \ , \\
        \\
        ~~\cdots \cdots\cdots  \ .    
        \end{cases}
    \end{aligned}
\end{equation}
By utilizing the $n$-th order perturbation of the Christoffel symbol, we can directly derive the expression for the $n$-th order perturbation of the Ricci tensor $R_{\mu\nu}$, and subsequently determine the expression for the $n$-th order perturbation of the Einstein tensor $G_{\mu\nu}$. The dominant contribution to induced gravitational waves often arises from source terms composed of scalar perturbations. Therefore, vertex diagrams formed by scalar perturbations are particularly important.  If we only consider the scalar perturbations $\psi$ and $\phi$, the $n$-th order perturbations of $g^{\mu\nu}$ can be written as
\begin{equation}
    \begin{aligned}
      \delta^n g^{\mu \nu} &=\delta^n g^{\mu \nu}_{ss}=\frac{2^n}{a^2}\left(\left( 
-1 \right)^{n+1}\phi^nn^{\mu}n^{\nu}+\psi^n m^{\mu\nu} \right) =\frac{2^n}{a^2}
\left(
  \begin{array}{ccc}
   \left( 
-1 \right)^{n+1} \phi^n  & 0  \\
    0  & \psi^n \delta^{ij} \\
  \end{array}
\right)  \ .
    \end{aligned}
\end{equation}
In this case, Eq.~(\ref{eq:Gamma}) can be simplified as
\begin{equation}\label{eq:Gammas}
    \begin{aligned}
        \delta^n \Gamma^{\sigma}_{\mu\nu}&=(-1)^{n}2^{n-1}\left[m_{\mu\nu} \left( n^{\sigma} \phi^{n-1} \left(2\mathcal{H}\phi+2\mathcal{H}\psi+\psi'  \right)+(-1)^n\psi^{n-1}\partial^{\sigma}\psi \right)  + n^{\sigma}\phi^{n-1}\left( n_{\nu}\partial_{\mu}\phi-n_{\mu}n_{\nu}\phi'+n_{\mu}\partial_{\nu}\phi  \right)   \right.\\
        &\left. + (-1)^n \psi^{n-1}  \left( n_{\mu}n_{\nu} \partial^{\sigma} \phi +m^{\sigma}_{\nu}n_{\mu}\psi'- m^{\sigma}_{\nu}\partial_{\mu}\psi+m^{\sigma}_{\mu}n_{\nu}\psi'- m^{\sigma}_{\mu}\partial_{\nu}\psi  \right) \right] \ .
    \end{aligned} 
\end{equation}
The $n$-th order perturbation of Ricci tensor is given by
\begin{equation}
    \begin{aligned}
        \delta^n R_{\mu\nu}&=\partial_{\sigma}\delta^n\Gamma^{\sigma}_{\mu\nu}-\partial_{\nu}\delta^n\Gamma^{\sigma}_{\sigma\mu}
        +\sum^{n}_{k=0}C^k_n\left(\delta^{n-k}\Gamma^{\sigma}_{\sigma\alpha}\delta^{k}\Gamma^{\alpha}_{\mu\nu}-\delta^{n-k}\Gamma^{\sigma}_{\mu\alpha}\delta^{k}\Gamma^{\alpha}_{\sigma\nu}\right) \\
        &=(-1)^{n}2^{n-1}\partial_{\sigma}\left[m_{\mu\nu} \left( n^{\sigma} \phi^{n-1} \left(2\mathcal{H}\phi+2\mathcal{H}\psi+\psi'  \right)+(-1)^n\psi^{n-1}\partial^{\sigma}\psi \right)  + n^{\sigma}\phi^{n-1}\left( n_{\nu}\partial_{\mu}\phi-n_{\mu}n_{\nu}\phi'+n_{\mu}\partial_{\nu}\phi  \right)   \right.\\
        &\left. + (-1)^n \psi^{n-1}  \left( n_{\mu}n_{\nu} \partial^{\sigma} \phi +m^{\sigma}_{\nu}n_{\mu}\psi'- m^{\sigma}_{\nu}\partial_{\mu}\psi+m^{\sigma}_{\mu}n_{\nu}\psi'- m^{\sigma}_{\mu}\partial_{\nu}\psi  \right) \right]\\
        &-2^{n-1} \partial_{\nu}\left[n_{\mu}\left((-1)^{n}\phi^{n-1}\phi'+3\psi^{n-1}\psi'  \right) +(-1)^{n-1}\phi^{n-1}\partial_{\mu}\phi-3\psi^{n-1}\partial_{\mu}\psi  \right]\\
        &+C^k_n\Bigg( (-1)^{-k+n}2^{-2+n} \phi^{-2-k}\psi^{-2-k}\left( -(-1)^k\phi^{k+n}\psi^{2+k}\phi'  \left( m_{\mu\nu} \left(  2\mathcal{H}\phi+2\mathcal{H}\psi +\psi' \right) +n_{\nu}\partial_{\nu}\phi  \right)\right.\\
&\left.- 3(-1)^{n}\phi^{1+2k}\psi^{1+n}  \left( m_{\mu\nu}\left( \psi^{'2}+2\mathcal{H}\phi\psi'+2\mathcal{H}\psi \psi'  \right)+n_{\nu}\psi'\partial_{\mu}\phi    \right)-3(-1)^{k+n}\phi^{2+k}\psi^{k+n} \right.\\
&\left.  \left(  m_{\mu\nu}\partial_{\alpha}\psi\partial^{\alpha}\psi+\partial_{\mu}\psi\left(  n_{\mu}\psi'-2\partial_{\mu}\psi \right)      \right) +(-1)^{2k}\phi^{1+n}\psi^{1+2k}\left(-m_{\mu\nu}\partial_{\alpha}\psi \partial^\alpha   \phi+\partial_{\mu}\psi\partial_{\nu}\phi+\partial_{\mu}\phi    n_{\nu}\phi^{'2}-\partial_{\mu}\phi\phi'\partial_{\mu}\phi\right)  \right.\\
&\left. +n_{\mu}\left( 3(-1)^n\phi^{1+2k}\psi^{1+n}\psi' n_{\nu}\phi'-\psi'\partial_{\nu}\phi  +(-1)^k \phi^{k+n}\psi^{2+k} \left( n_{\nu}\phi^{'2}-\phi'\partial_{\nu}\phi  \right)\right.\right.\\
&\left.\left.-(-1)^{2k}\phi^{1+n}\psi^{1+2k}  \left( n_{\nu}\partial_{\alpha}\phi\partial^{\alpha}\phi+\psi'\partial_{\nu}\phi   \right)-3(-1)^{k+n}\phi^{k+n}\psi^{k+n}\left(n_{\nu}\partial_{\alpha}\psi\partial^{\alpha}\phi+\psi'\partial_{\nu}\psi     \right) \right)  \right)\\
& (-1)^{-k+n} 2^{-2+n}\phi^{-1-k}\psi^{-1-k}  \left((-1)^{k+1}\phi^{n}\psi^{1+k}\left((-1)^{k}\psi^4 \left(m_{\mu \nu}\left(\psi^{'2}+2 \mathcal{H}\left(\phi+\psi\right)\psi'\right)\right.\right.\right.\\
&\left.\left.\left.+m_{\mu \alpha} n_\nu\left(2 \mathcal{H}\left(\phi+\psi\right)+  \psi'\right) \partial^{\alpha} \phi\right)   +\phi^4 \partial_\mu  \phi\left(n_\nu\phi'-\partial_\nu  \phi\right)\right)\right. \\
&+n_\mu\left(-(-1)^{k} \phi^{n} \psi^{1+k}\left( \phi^4\left(-n_\nu \phi^{'2}+ \phi' \partial_\nu  \phi\right)+(-1)^{k} \psi^4\left(n_\nu \partial_\alpha  \phi \partial^{\alpha} \phi+ \psi' \partial_\nu  \phi\right)\right)+\right. \\
& \left.(-1)^{n} \phi^{1+k} \psi^{n}\left(- \phi^4\left(m_{\nu \alpha}\left(2 \mathcal{H}\left( \phi+ \psi\right)+  \psi'\right)+n_{\nu} \partial_\alpha  \phi\right) \partial^{\alpha} \phi+(-1)^{k} \psi^4\left(3 n_\nu \psi^{'2}+ \psi'\left(m_{\nu \alpha} \partial^{\alpha} \psi-4 \partial_\nu  \psi\right)\right)\right)\right) \\
&+(-1)^{n}\phi^{1+k}\psi^{n} \left(-\phi^4\left(m_{\mu \nu}\left(  \psi^{'2}+2 \mathcal{H}\phi\psi'+2 \mathcal{H}\psi\psi'\right)+n_\nu\psi' \partial_\mu  \phi\right)+\right. \\
& \left.\left.(-1)^{k}\psi^4\left(-2 m_{\mu \nu} \partial_\alpha  \psi \partial^{\alpha } \psi-\partial_\mu  \psi\left(4 n_\nu\psi'+m_{\nu \alpha} \partial^{\alpha } \psi-6 \partial_\nu  \psi\right)+m_{\mu \alpha} \partial^{\alpha } \psi\left(n_\nu\psi'-\partial_\nu  \psi+m_{\nu \sigma} \partial^{\sigma } \psi\right)\right)\right) \right)
  \Bigg) \ .
    \end{aligned}
\end{equation}
The $n$-th order perturbation of the Einstein tensor can be expressed as follows
\begin{equation}
    \begin{aligned}
        \delta^n G_{\mu\nu}=\delta^n R_{\mu\nu}-\frac{1}{2}\left(g^{(0)}_{\mu\nu} \delta R + \delta g_{\mu\nu} \delta^{n-1} R \right) \ ,
    \end{aligned}
\end{equation}
where
\begin{equation}
    \begin{aligned}
        \delta^n R=\delta^n \left( g^{\mu\nu} R_{\mu\nu}  \right)=\sum^{n}_{k=0}C^k_n~\frac{2^{n-k}}{a^2}\left(\left( 
-1 \right)^{{n-k}+1}\phi^{n-k}n^{\mu}n^{\nu}+\psi^{n-k} m^{\mu\nu} \right)~  \delta^k R_{\mu\nu}  \ .
    \end{aligned}
\end{equation}
During the radiation-dominated era, we set $\delta^n p=\frac{1}{3}\delta^n \rho$. The energy-momentum tensor of perfect fluid can be written as
\begin{eqnarray}\label{eq:T0}
    T_{\mu\nu}=\frac{4}{3}\rho^{} u^{}_\mu u^{}_\nu + \frac{1}{3} \rho^{} g_{\mu\nu} \ .
\end{eqnarray} 
The $n$-th order perturbation of $T_{\mu\nu}$ is gievn by
\begin{equation}
    \begin{aligned}
        \delta^n T_{\mu\nu}=\frac{4}{3}\sum_{m=0}^{n}\sum^{m}_{l=0} C_n^m\delta^{n-m} \rho C_m^l
 \delta^{m-l}u_{\mu}  \delta^{l}u_{\nu}+\frac{1}{3}\sum^{n}_{m=0} C_n^m\delta^{n-m}g_{\mu\nu}  \delta^{m}\rho \ .
    \end{aligned}
\end{equation}
By utilizing the time-time component and the time-space component  of the $n$-th order perturbations of Einstein's field equation, we can represent density perturbations $\delta^n \rho$ and velocity perturbations $\delta^n u_{\mu}$ as metric perturbations. Substituting the calculated density and velocity perturbations into the space-space component of the $n$-th order perturbations of Einstein's field equations yields the source terms associated with various vertex diagrams.

\section{Two-branch vertex}\label{sec:B}
We present the specific expressions for the source terms corresponding to all the 2-branch vertices of the induced gravitational waves.  In the vertex diagrams, the dashed lines, wavy lines and spring-like lines represent scalar, vector, and tensor perturbations, respectively.

$(1)$ Induced gravitational waves sourced by lower-order scalar perturbations $\phi$ and $\psi$.
\begin{equation}\label{eq:Sss}
    \begin{aligned}
\mathcal{S}_{\phi\phi,ij}+\mathcal{S}_{\psi\phi,ij}+\mathcal{S}_{\psi\psi,ij}&=\partial_i  \psi \partial_j  \phi+\frac{\partial_i \psi^{\prime} \partial_j  \phi}{\mathcal{H}}+\partial_i  \phi \partial_j  \psi-3 \partial_i  \psi \partial_j  \psi+\frac{\partial_i  \phi \partial_j  \psi}{\mathcal{H}}+\frac{\partial_i { }^{\prime} \psi \partial_j { }^{\prime} \psi}{\mathcal{H}^2}\\
&-2 \phi \partial_j \partial_i  \phi-2 \psi\partial_j \partial_i  \psi \ .
    \end{aligned}
\end{equation}

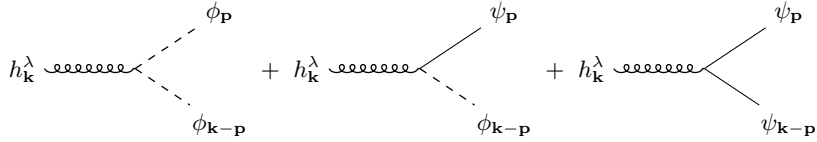
\begin{figure}[htbp]
    \centering
\begin{equation*}
    \begin{tikzpicture}[baseline=(hk.base)]
        \begin{feynhand}
        \vertex (hk) at (0,0) {$h^{\lambda}_{\mathbf{k}}$};
        \vertex (center) at (1.5,0);
        \vertex (phi) at (2.625,0.75) {$\phi_\mathbf{p}$};
        \vertex (psi) at (2.625,-0.75) {$\phi_{\mathbf{k}-\mathbf{p}}$};
        \propag [gluon] (hk) to (center);
        \propag [sca] (center) to  (phi);
        \propag [sca] (center) to  (psi);
        \end{feynhand}
    \end{tikzpicture}        
    +
    \begin{tikzpicture}[baseline=(hk.base)]
        \begin{feynhand}
        \vertex (hk) at (0,0) {$h^{\lambda}_{\mathbf{k}}$};
        \vertex (center) at (1.5,0);
        \vertex (phi) at (2.625,0.75) {$\psi_\mathbf{p}$};
        \vertex (psi) at (2.625,-0.75) {$\phi_{\mathbf{k}-\mathbf{p}}$};
        \propag [gluon] (hk) to (center);
        \propag (center) to  (phi);
        \propag [sca] (center) to  (psi);
        \end{feynhand}
    \end{tikzpicture} 
    +
    \begin{tikzpicture}[baseline=(hk.base)]
        \begin{feynhand}
        \vertex (hk) at (0,0) {$h^{\lambda}_{\mathbf{k}}$};
        \vertex (center) at (1.5,0);
        \vertex (phi) at (2.625,0.75) {$\psi_\mathbf{p}$};
        \vertex (psi) at (2.625,-0.75) {$\psi_{\mathbf{k}-\mathbf{p}}$};
        \propag [gluon] (hk) to (center);
        \propag (center) to  (phi);
        \propag (center) to  (psi);
        \end{feynhand}
    \end{tikzpicture} 
\end{equation*}
\caption{\label{fig_supp:1} The 2-branch vertex diagram of higher-order gravitational waves induced by the product of two lower-order scalar perturbations.}
\end{figure}
For example, if we need to calculate the source term of the 10th-order gravitational waves induced by the product of two 5th-order scalar perturbations, we only need to make the following substitution in the source term of Eq.~(\ref{eq:Sss}): $\phi\to \frac{1}{5!}\phi^{(5)} \ , \ \psi\to \frac{1}{5!}\psi^{(5)}$.

\

$(2)$ Induced gravitational waves sourced by lower-order vector perturbation $V_i$.

\begin{equation}
    \begin{aligned}
\mathcal{S}_{VV,ij}&=-\frac{1}{2}V^{b} \partial_b \partial_i  V_{j}-\frac{1}{2}V^{b} \partial_b \partial_{j}  V_i+\frac{1}{2} \partial_b  V_{j} \partial^{b} V_i+\frac{ \partial_b\partial^b  V_i  \partial_c\partial^c V_{j}}{16\mathcal{H}^2}+\frac{1}{2} \partial_i  V^b \partial_{j } V_b+V^{b} \partial_j \partial_i  V_b \ .
    \end{aligned}
\end{equation}

\begin{figure}[htbp]
    \centering
    \begin{tikzpicture}[baseline=(hk.base)]
    \begin{feynhand}
        \vertex (hk) at (0,0) {$h^{\lambda}_{\mathbf{k}}$};
        \vertex (center) at (1.5,0);
        \vertex (phi) at (2.625,0.75) {$V^{\lambda_1}_\mathbf{p}$};
        \vertex (psi) at (2.625,-0.75) {$V^{\lambda_2}_{\mathbf{k}-\mathbf{p}}$};
        \propag [gluon] (hk) to (center);
        \propag [boson] (center) to  (phi);
        \propag [boson] (center) to  (psi);
    \end{feynhand}
\end{tikzpicture}  
\caption{\label{fig_supp:2} The 2-branch vertex diagram of higher-order gravitational waves induced by the product of two lower-order vector perturbations.}
\end{figure}

\

$(3)$ Induced gravitational waves sourced by lower-order tensor perturbation $h_{ij}$.

\begin{equation}
    \begin{aligned}
\mathcal{S}_{hh,ij}&=\frac{1}{2}   h_{i}^{b'}  h^{\prime}_{j b}-\frac{1}{2}   h^{b c}\partial_c \partial_b   h_{i j}+\frac{1}{2}  h^{b c} \partial_c \partial_i   h_{j b}+\frac{1}{2}  h^{b c} \partial_c \partial_j   h_{i b}+\frac{1}{2} \partial_b   h_{j c} \partial^{c}  h_{i}^b-\frac{1}{2} \partial_c   h_{j b} \partial^{c}  h_{i}^{b} \\
&-\frac{1}{4} \partial_i   h^{b c} \partial_j   h_{b c}-\frac{1}{2}   h^{b c}\partial_j \partial_i   h_{b c} \ .
    \end{aligned}
\end{equation}

\begin{figure}[htbp]
    \centering
    \begin{tikzpicture}[baseline=(hk.base)]
    \begin{feynhand}
        \vertex (hk) at (0,0) {$h^{\lambda}_{\mathbf{k}}$};
        \vertex (center) at (1.5,0);
        \vertex (phi) at (2.625,0.75) {$h^{\lambda_1}_\mathbf{p}$};
        \vertex (psi) at (2.625,-0.75) {$h^{\lambda_2}_{\mathbf{k}-\mathbf{p}}$};
        \propag [gluon] (hk) to (center);
        \propag [gluon] (center) to  (phi);
        \propag [gluon] (center) to  (psi);
    \end{feynhand}
\end{tikzpicture}   
\caption{\label{fig_supp:3} The 2-branch vertex diagram of higher-order gravitational waves induced by the product of two lower-order tensor perturbations.}
\end{figure}

\

$(4)$ Induced gravitational waves sourced by lower-order tensor perturbation $h_{ij}$  and lower-order scalar perturbations $\phi$ and $\psi$.

\begin{equation}
    \begin{aligned}
\mathcal{S}_{h\phi,ij}+\mathcal{S}_{h\psi,ij}&=   h_{ i j}^{\prime \prime} \phi+2  h_{ i j}^{\prime} \mathcal{H}  \phi+\frac{1}{2}   h_{ i j}^{\prime} \phi'-2   h_{ i j} \mathcal{H}  \phi'-\frac{1}{2}   h_{ i j}^{\prime} \psi'-8   h_{ i j} \mathcal{H} \psi'-3  h_{ i j} \psi''+  \psi \partial_b \partial^b   h_{ i j} \\
&-h_{ i j} \partial_b \partial^{b } \phi+\frac{8}{3}  h_{ i j} \partial_b \partial^{b} \psi- h_v^b \partial_b \partial_ i \psi-   h_{ i }^b \partial_b \partial_{j}  \psi+\frac{1}{2} \partial_b   h_{ i j} \partial^{b} \phi+\frac{3}{2} \partial_b   h_{ i j} \partial^{b} \psi-\frac{1}{2} \partial^{b} \phi \partial_ i   h_{j b} \\
&-\frac{1}{2} \partial^{b} \psi \partial_ i   h_{j b}- \frac{1}{2} \partial^{b} \phi \partial_j   h_{ i b}-\frac{1}{2} \partial^{b} \psi \partial_j   h_{ i b} \ .
    \end{aligned}
\end{equation}

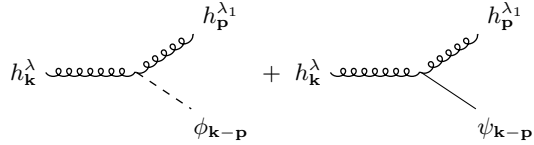
\begin{figure}[htbp]
    \centering
\begin{equation*}
    \begin{tikzpicture}[baseline=(hk.base)]
        \begin{feynhand}
        \vertex (hk) at (0,0) {$h^{\lambda}_{\mathbf{k}}$};
        \vertex (center) at (1.5,0);
        \vertex (phi) at (2.625,0.75) {$h^{\lambda_1}_\mathbf{p}$};
        \vertex (psi) at (2.625,-0.75) {$\phi_{\mathbf{k}-\mathbf{p}}$};
        \propag [gluon] (hk) to (center);
        \propag [gluon] (center) to  (phi);
        \propag [sca] (center) to  (psi);
        \end{feynhand}
    \end{tikzpicture}        
    +
    \begin{tikzpicture}[baseline=(hk.base)]
        \begin{feynhand}
        \vertex (hk) at (0,0) {$h^{\lambda}_{\mathbf{k}}$};
        \vertex (center) at (1.5,0);
        \vertex (phi) at (2.625,0.75) {$h^{\lambda_1}_\mathbf{p}$};
        \vertex (psi) at (2.625,-0.75) {$\psi_{\mathbf{k}-\mathbf{p}}$};
        \propag [gluon] (hk) to (center);
        \propag [gluon] (center) to  (phi);
        \propag  (center) to  (psi);
        \end{feynhand}
    \end{tikzpicture} 
\end{equation*}
\caption{\label{fig_supp:4} The 2-branch vertex diagram of higher-order gravitational waves induced by the product of lower-order tensor perturbations and lower-order scalar perturbations.}
\end{figure}

\

$(5)$ Induced gravitational waves sourced by lower-order vector perturbation $V_{i}$  and lower-order scalar perturbations $\phi$ and $\psi$.
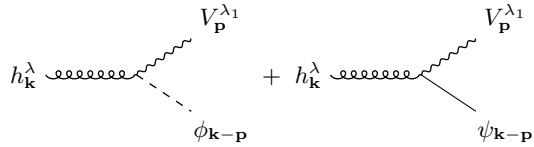
\begin{figure}[htbp]
    \centering
\begin{equation*}
    \begin{tikzpicture}[baseline=(hk.base)]
        \begin{feynhand}
        \vertex (hk) at (0,0) {$h^{\lambda}_{\mathbf{k}}$};
        \vertex (center) at (1.5,0);
        \vertex (phi) at (2.625,0.75) {$V^{\lambda_1}_\mathbf{p}$};
        \vertex (psi) at (2.625,-0.75) {$\phi_{\mathbf{k}-\mathbf{p}}$};
        \propag [gluon] (hk) to (center);
        \propag [boson] (center) to  (phi);
        \propag [sca] (center) to  (psi);
        \end{feynhand}
    \end{tikzpicture}        
    +
    \begin{tikzpicture}[baseline=(hk.base)]
        \begin{feynhand}
        \vertex (hk) at (0,0) {$h^{\lambda}_{\mathbf{k}}$};
        \vertex (center) at (1.5,0);
        \vertex (phi) at (2.625,0.75) {$V^{\lambda_1}_\mathbf{p}$};
        \vertex (psi) at (2.625,-0.75) {$\psi_{\mathbf{k}-\mathbf{p}}$};
        \propag [gluon] (hk) to (center);
        \propag [boson] (center) to  (phi);
        \propag  (center) to  (psi);
        \end{feynhand}
    \end{tikzpicture} 
\end{equation*}
\caption{\label{fig_supp:5} The 2-branch vertex diagram of higher-order gravitational waves induced by the product of lower-order vector perturbations and lower-order scalar perturbations.}
\end{figure}

\begin{equation}
    \begin{aligned}
\mathcal{S}_{V\phi,ij}+\mathcal{S}_{V\psi,ij}&=-2 \mathcal{H} \phi \partial_i  V_{j}-\frac{1}{2}  \phi' \partial_i  V_{j}-\frac{1}{2} \psi' \partial_i  V_{j}-  \phi \partial_i  V^{\prime}_{j}-\frac{\partial_b \partial^b  V_{j} \partial_i  \phi}{4\mathcal{H}}+  V_{j}^{\prime} \partial_i  \psi+ 2  V_{j} \mathcal{H} \partial_i  \psi+  V_{j} \partial_i  \psi^{\prime}\\
&-\frac{\partial_b \partial^b  V_{j} \partial_i  \psi'}{4\mathcal{H}^2}-2 \mathcal{H} \phi \partial_i  V_{j}-\frac{1}{2}  \phi' \partial_i  V_{j}-\frac{1}{2} \psi' \partial_i  V_{j}-  \phi \partial_i  V^{\prime}_{j}-\frac{\partial_b \partial^b  V_{j} \partial_i  \phi}{4\mathcal{H}}+  V_{j}^{\prime} \partial_i  \psi+ 2  V_{j} \mathcal{H} \partial_i  \psi\\
&+  V_{j} \partial_i  \psi^{\prime}-\frac{\partial_b \partial^b  V_{j} \partial_i  \psi'}{4\mathcal{H}^2} \ .
    \end{aligned}
\end{equation}

\

$(6)$ Induced gravitational waves sourced by lower-order vector perturbation $V_{i}$  and lower-order tensor perturbation $h_{ij}$.

\begin{equation}
    \begin{aligned}
\mathcal{S}_{Vh,ij}&=\frac{1}{2}  V^{b\prime} \partial_b  h_{ i  j}+ V^b \mathcal{H} \partial_b  h_{ i  j}+  V^b \partial_b  h_{ i  j}^{\prime}-\frac{1}{2}  h_{ j b}^{\prime} \partial^b  V_ i-\frac{1}{2} h_{ i b}^{\prime} \partial^b  V_{ j}-\frac{1}{2} V^{b\prime} \partial_ i  h_{ j b}-  V^b \mathcal{H} \partial_ i  h_{ j b}-\frac{1}{2}  V^b \partial_ i  h_{ j b}^{\prime} \\
& -\frac{1}{2}  V^{b\prime} \partial_ j  h_{ i b}-  V^b \mathcal{H} \partial_ j  h_{ i b}-\frac{1}{2}  V^b \partial_ j  h_{ i b}^{\prime} \ .
    \end{aligned}
\end{equation}

\begin{figure}[htbp]
    \centering
    \begin{tikzpicture}[baseline=(hk.base)]
    \begin{feynhand}
        \vertex (hk) at (0,0) {$h^{\lambda}_{\mathbf{k}}$};
        \vertex (center) at (1.5,0);
        \vertex (phi) at (2.625,0.75) {$h^{\lambda_1}_\mathbf{p}$};
        \vertex (psi) at (2.625,-0.75) {$V^{\lambda_2}_{\mathbf{k}-\mathbf{p}}$};
        \propag [gluon] (hk) to (center);
        \propag [gluon] (center) to  (phi);
        \propag [boson] (center) to  (psi);
    \end{feynhand}
\end{tikzpicture}   
\caption{\label{fig_supp:6} The 2-branch vertex diagram of higher-order gravitational waves induced by the product of lower-order tensor perturbations and  lower-order vector perturbations.}
\end{figure}

\section{Third-order tensor-scalar induced gravitational waves}\label{sec:C}

The explicit expression of $h^{\lambda,(3)}_{\mathbf{k},s h_{s h}}$ is given by
\begin{equation}\label{eq:h30}
\begin{aligned}
h_{s h_{s h}}^{\lambda,(3)}(\eta, \mathbf{k})&=\frac{16}{3}\int \frac{d^3 p}{(2 \pi)^{3 / 2}} \int \frac{d^3 q}{(2 \pi)^{3 / 2}} \varepsilon^{\lambda, l m}(\mathbf{k}) \varepsilon_{l m}^{\lambda_1}(\mathbf{p})\varepsilon^{\lambda_1,ij}\left(\mathbf{p}\right) \varepsilon^{\lambda_2}_{ij}\left(\mathbf{q}\right)v^2  \\
&~~\times\int_0^{\bar{x}} d\bar{x} 
\frac{\bar{x}}{x}\sin\left( x-\bar{x} \right) f^{(3)}_{s h_{s h}}\left(u,v,\bar{u},\bar{v},\bar{x}  \right) \zeta_{\mathbf{k}-\mathbf{p}}\zeta_{\mathbf{p}-\mathbf{q}}\mathbf{h}^{\lambda_2}_{\mathbf{q}} \ .
\end{aligned}
\end{equation}
In Eq.~(\ref{eq:h30}), $f^{(3)}_{s h_{sh}}\left(u,v,\bar{u},\bar{v},x  \right)$  can be represented as
\begin{equation}
\begin{aligned}
f^{(3)}_{s h_{s h}}\left(u,v,\bar{u},\bar{v},x  \right)=&-2 T_{\phi}\left(ux\right)f^{(2)}_{sh}\left( \bar{u},\bar{v},y \right)+T_{\phi}\left(ux\right)p^2 I^{(2)}_{sh}\left( \bar{u},\bar{v},y \right)-\frac{u}{v^2x}p^2I^{(2)}_{sh}\left( \bar{u},\bar{v},y \right)\frac{d}{d(ux)}T_{\phi}\left(ux \right)  \\
&  +\frac{u^2}{3v^2} T_{\phi}\left(ux \right)p^2I^{(2)}_{sh}\left( \bar{u},\bar{v},y \right) +\frac{\left(1-v^2-u^2 \right)}{2v^2} T_{\phi}\left(ux \right) p^2I^{(2)}_{sh}\left( \bar{u},\bar{v},y \right) \ ,
\end{aligned}
\end{equation}
where
\begin{equation}
\begin{aligned}
f_{sh}^{(2)}\left( \bar{u},\bar{v}, y \right)&=\frac{10u}{x}\frac{d}{d(ux)} T_{\phi}\left( ux \right)T_{h}\left( vx \right)+3u^2\frac{d^2}{d(ux)^2}T_{\phi}\left(ux \right)T_{h}\left( vx \right)+\frac{5}{3}u^2 T_{\phi}\left( ux \right)T_{h}\left( vx \right) \nonumber\\
	&+\left(1-v^2-u^2\right)T_{\phi}\left( ux \right)T_{h}\left( vx \right)+2v^2T_{\phi}\left( ux \right)T_{h}\left( vx \right)  \ ,  \\
 p^2I^{(2)}_{sh}\left( \bar{u},\bar{v},y \right)&=4 \int_{0}^{y} d\bar{y} \left( \frac{\bar{y}}{y}\sin\left( y-\bar{y} \right) f^{(2)}_{sh}\left( \bar{u},\bar{v},\bar{y} \right)  \right) \  , \\
 T_{\phi}(x)&=\frac{9}{x^{2}}\left(\frac{\sqrt{3}}{x} \sin \left(\frac{x}{\sqrt{3}}\right)-\cos \left(\frac{x}{\sqrt{3}}\right)\right) \ .
\end{aligned}
\end{equation}
We have defined $|\mathbf{k}-\mathbf{p}|=u|\mathbf{k}|$, $|\mathbf{p}|=v|\mathbf{k}|$,  $|\mathbf{p}-\mathbf{q}|=\bar{u}|\mathbf{p}|$,  $x=|\mathbf{k}|\eta$ , and $y=|\mathbf{p}|\eta$.

\bibliography{vertex}
\end{document}